\documentclass[a4paper,12pt]{article}
\pdfoutput=1
\usepackage{feynmp-auto,expdlist}
\usepackage{amsmath, amsfonts, amssymb}
\usepackage{graphicx}
\usepackage{enumerate}
\usepackage{hyperref}
\usepackage{latexsym}
\usepackage{dsfont}
\usepackage{hepnicenames}
\usepackage{enumerate}
\usepackage{soul}
\usepackage[normalem]{ulem}
\usepackage{bbold}
\usepackage{units}

\newcommand{\A}{{\cal A}}
\newcommand{\B}{{\cal B}}

\newcommand{\G}{{\cal G}}

\newcommand{\W}{{\cal W}}
\newcommand{\Z}{{\cal Z}}
\newcommand{\M}{{\cal M}}
\renewcommand{\H}{{\cal H}}
\newcommand{\X}{{\cal X}}
\renewcommand{\S}{{\cal S}}
\usepackage{mathrsfs,graphicx,rotating,amsmath,amsfonts,mathtools,booktabs,amssymb,wasysym}
\usepackage{hyperref}\usepackage{slashed}
\usepackage[nosort]{cite}
\usepackage[table,xcdraw,dvipsnames]{xcolor}
\usepackage{bm}
\usepackage{graphicx}
\usepackage{multirow,multicol}
\hypersetup{colorlinks,bookmarksopen,bookmarksnumbered,
linkcolor=blus,pdfstartview=FitH,urlcolor=rossos,citecolor=verde}
\allowdisplaybreaks

\newcommand{\mDG}{M_{\scriptscriptstyle{\rm DG}}}

\newcommand{\NDC}{\ndc}
\renewcommand{\[}{\left[}
\renewcommand{\]}{\right]}
\renewcommand{\(}{\left(}
\renewcommand{\)}{\right)}

\def\Lag{\mathscr{L}}

\newcommand{\mio}[1]{}

 \newcommand{\med}[1]{\langle #1\rangle}

\def\bpm{\begin{pmatrix}}
\def\epm{\end{pmatrix}}

\usepackage{mathrsfs}

 \newcommand{\fig}[1]{~\ref{fig:#1}}
\newcommand{\sfrac}[2]{#1/#2}
 \newcommand{\One}{1\!\!\hbox{I}}

\renewcommand{\Re}{{\rm Re}\,}
\renewcommand{\Im}{{\rm Im}\,}
\allowdisplaybreaks
\usepackage{multicol}
\usepackage{color}
\definecolor{rosso}{cmyk}{0,1,1,0.4}
\definecolor{rossos}{cmyk}{0,1,1,0.55}
\definecolor{rossoc}{cmyk}{0,1,1,0.2}
\definecolor{blu}{cmyk}{1,1,0,0.3}
\definecolor{blus}{cmyk}{1,1,0,0.6}
\definecolor{bluc}{cmyk}{1,1,0,0.1}
\definecolor{verde}{cmyk}{0.92,0,0.59,0.25}
\definecolor{verdec}{cmyk}{0.92,0,0.59,0.15}
\definecolor{verdes}{cmyk}{0.92,0,0.59,0.4}

\newcommand{\eq}[1]{~{\rm (\ref{eq:#1})}}

\newcommand{\GeV}{\,{\rm GeV}}
\newcommand{\TeV}{\,{\rm TeV}}
\newcommand{\cm}{\,{\rm cm}}

\newcommand{\Tr}{\,{\rm Tr}}
\newcommand{\diag}{\,{\rm diag}}

\def\circa#1{\,\raise.3ex\hbox{$#1$\kern-.75em\lower1ex\hbox{$\sim$}}\,}

\newcommand{\beq}{\begin{equation}}
\newcommand{\eeq}{\end{equation}}

\newcommand{\bea}{\begin{eqnarray}}
\newcommand{\eea}{\end{eqnarray}}
\newcommand{\be}{\begin{equation}}
\newcommand{\ee}{\end{equation}}
\font\tenrsfs=rsfs10 at 12pt
\font\sevenrsfs=rsfs7
\font\fiversfs=rsfs5
\newfam\rsfsfam
\textfont\rsfsfam=\tenrsfs
\scriptfont\rsfsfam=\sevenrsfs
\scriptscriptfont\rsfsfam=\fiversfs

\newcommand{\D}{{\cal D}}

\newcommand{\La}{\mathscr{L}}

\newsavebox\MBox

\newcommand{\Sp}{\,{\rm Sp}}

\newcommand{\LDC}{\Lambda_{\rm DC}}

\newcommand{\gDC}{g_{{\rm DC}}}
\newcommand{\adc}{\alpha_{\scriptscriptstyle{\rm DC}}}

\newcommand{\SU}{\,{\rm SU}}
\newcommand{\SO}{\,{\rm SO}}
\newcommand{\U}{\,{\rm U}}

\def\circa#1{\,\raise.3ex\hbox{$#1$\kern-.75em\lower1ex\hbox{$\sim$}}\,}
\makeatletter

\font\ital=cmu10

\def\hhref#1{\href{http://arxiv.org/abs/#1}{arXiv:#1}}
\usepackage{xstring}
\newcommand{\hhrefq}[1]{\IfSubStr{#1}{:}{\href{http://inspirehep.net/search?ln=en&ln=en&p=#1&of=hb&action_search=Search&sf=&so=d&rm=&rg=25&sc=0}{InSpire:#1}}{\hhref{#1}}}

\def\art{\@ifnextchar[{\eart}{\oart}}
\def\eart[#1]#2#3#4#5#6{{\rm #2}, {\em #3 \bf #4} {\rm (#6) #5} ({\em #1})}
\def\article{\@ifnextchar[{\earticle}{\oarticle}}
\def\oarticle#1#2#3#4#5#6{{\rm #1}, {\ital ``#6''}, {\rm #2 #3 (#5) #4}}
\def\earticle[#1]#2#3#4#5#6#7{{\rm #2}, {\ital ``#7''}, {\rm #3 #4 (#6) #5}  [\hhrefq{#1}]}
\def\hepart[#1]#2{{\rm #2, \sl#1}}
\def\heparticle[#1]#2#3{#2, {\ital ``#3''} [\hhrefq{#1}]}
\newcommand{\doi}[1]{\href{http://dx.doi.org/#1}{[link]}}

\newcommand{\hhrefqq}[1]{\IfBeginWith{#1}{10.}{\href{https://doi.org/#1}{doi:#1}}{\hhrefq{#1}}}
\def\earticle[#1]#2#3#4#5#6#7{{\rm #2}, {\ital ``#7''}, {\rm #3 #4 (#6) #5}  [\hhrefqq{#1}]}

\renewenvironment{thebibliography}[1]
     {\begin{multicols}{2}[\section*{\refname}]%
      \@mkboth{\MakeUppercase\refname}{\MakeUppercase\refname}%
      \list{\@biblabel{\@arabic\c@enumiv}}%
           {\settowidth\labelwidth{\@biblabel{#1}}%
            \leftmargin\labelwidth
            \advance\leftmargin\labelsep
            \@openbib@code
            \usecounter{enumiv}%
            \let\p@enumiv\@empty
            \renewcommand\theenumiv{\@arabic\c@enumiv}}%
      \sloppy
      \clubpenalty4000
      \@clubpenalty \clubpenalty
      \widowpenalty4000%
      \sfcode`\.\@m}
     {\def\@noitemerr
       {\@latex@warning{Empty `thebibliography' environment}}%
      \endlist\end{multicols}}

\newcounter{alphaequation}[equation]
\def\thealphaequation{\theequation\hbox to
0.6em{\hfil\alph{alphaequation}\hfil}}
\def\eqnsystem#1{
\def\@eqnnum{{\rm (\thealphaequation)}}
\def\@@eqncr{\let\@tempa\relax \ifcase\@eqcnt \def\@tempa{& & &} \or
  \def\@tempa{& &}\or \def\@tempa{&}\fi\@tempa
  \if@eqnsw\@eqnnum\refstepcounter{alphaequation}\fi
\global\@eqnswtrue\global\@eqcnt=0\cr}
\refstepcounter{equation} \let\@currentlabel\theequation \def\@tempb{#1}
\ifx\@tempb\empty\else\label{#1}\fi
\refstepcounter{alphaequation}
\let\@currentlabel\thealphaequation
\global\@eqnswtrue\global\@eqcnt=0 \tabskip\@centering\let\\=\@eqncr
$$\halign to \displaywidth\bgroup \@eqnsel\hskip\@centering
$\displaystyle\tabskip\z@{##}$&\global\@eqcnt\@ne
\hskip2\arraycolsep\hfil${##}$\hfil& \global\@eqcnt\tw@\hskip2\arraycolsep
$\displaystyle\tabskip\z@{##}$\hfil
\tabskip\@centering&\llap{##}\tabskip\z@\cr}
\def\endeqnsystem{\@@eqncr\egroup$$\global\@ignoretrue} \makeatother

\newcommand{\ndc}{N_{\rm DC}}
\renewcommand{\ndc}{{\cal N}}

\definecolor{Gray}{gray}{0.95}

\def\bal#1\eal{\begin{align}#1\end{align}}

\begin{document}
\vspace{1.5cm}

\begin{center}
{\Large\LARGE\Huge \bf \color{rossos}
Dark Matter from self-dual\\[4mm] gauge/Higgs dynamics}\\[1cm]
{\bf Dario Buttazzo$^{a}$, Luca Di Luzio$^{a,b}$, Giacomo Landini$^{a,b}$, \\
Alessandro Strumia$^{b}$, Daniele Teresi$^{a,b}$}\\[7mm]

{\it $^a$ INFN, Sezione di Pisa, Largo Bruno Pontecorvo 3, I-56127 Pisa, Italy}\\[1mm]
{\it $^b$ Dipartimento di Fisica ``E. Fermi'', Universit\`a di Pisa, Italy
}\\[1mm]

\vspace{0.5cm}

\begin{quote}\large
We show that a new gauge group with one new scalar
leads to automatically stable Dark Matter candidates.
We consider theories where the Higgs phase is dual to the confined phase:
it is known that SU(2) gauge theories with a scalar doublet (like the Standard Model)
obey this non-trivial feature.
We provide a general criterion, showing that this self-duality holds
for SU($N$), SO($N$), Sp($N$) and $G_2$ gauge dynamics with a scalar field in the fundamental representation.
The resulting Dark Matter phenomenology has non-trivial features that are characteristic of the group, and that we discuss case by case.
Just to mention a few, SU($N$) has an accidental conserved dark baryon number
and charge conjugation, 
SO($2N+1$) leads to stable glue-balls thanks to a special parity,
$G_2$ leads to a Dark Matter system analogous to neutral kaons.
The cosmological Dark Matter abundance is often reproduced for masses around 100 TeV:
all constraints are satisfied and lighter dark glue-balls can affect Higgs physics. 
These theories acquire additional interest and predictivity assuming that
both the dark and weak scales are dynamically generated. 
\end{quote}

\thispagestyle{empty}
\bigskip

\end{center}

\setcounter{footnote}{0}

\newpage
\tableofcontents

\newpage

\section{Introduction}
We know that Dark Matter (DM) exists because we observed its collective
gravitational interactions,
but we do not know what DM is.
Many theories are possible.
Since gauge
interactions are maximally predictive in relativistic quantum field theory,
it makes sense to explore theories where gauge dynamics leads to DM.
We thereby add a new `dark' gauge group $\G$.
Its glue-balls could be DM without any interaction with the Standard Model  sector.
In order to thermally reproduce the cosmological DM abundance
we  minimally connect the dark sector to the Standard Model
by adding one scalar field $\S$ charged under $\G$.
Depending on $\G$, this leads to non-trivial accidental symmetries that imply 
DM stability with non-standard physics.
Despite that light elementary scalars are considered as unnatural by some theorists,
interesting DM matter models based on scalars have already been  proposed:
\begin{itemize}
\item[1)] The most minimal DM model in terms of new degrees of freedom involves just one
singlet scalar $\S$~\cite{ScalarSinglet}.
This is stable imposing an ad-hoc $\mathbb{Z}_2$ symmetry $\S\to -\S$ and assuming that the $\S$ vacuum expectation value vanishes.
Direct detection bounds excluded a significant part of the parameter space of this model~\cite{ScalarSinglet}.

\item[2)] Next, if the field $\S$ is complex, describing two scalar degrees of freedom, it can be charged under a new $\G={\rm U}(1)$ gauge group.
A vacuum expectation value of $\S$ breaks U(1) to nothing and the resulting massive vector $A_\mu$ is a DM candidate,
stable thanks to charge conjugation, $\S\to \S^*$ and
$A_\mu\to - A_\mu$, which is a symmetry
if the U(1) has vanishing kinetic mixing with hypercharge~\cite{1111.4482}.

\item[3)]  A more interesting model where DM stability is automatically implied by the particle content
has been proposed in~\cite{0811.0172,0907.1007,1306.2329}, assuming
that the scalar $\S$ fills the fundamental representation 2
of a new $\SU(2) $ gauge group.
A vacuum expectation value of $\S$ breaks $\SU(2)$ to nothing and
the DM candidates are the three $\SU(2)$ vectors, 
which acquire a common mass because of an accidental custodial symmetry.
\end{itemize}
The SU(2) model admits two apparently different phases:  {\it Higgs} and {\it confined}. 
A non-trivial feature of the SU(2) model --- interesting even from a purely theoretical point of view ---
is that the two phases give the same spectrum of asymptotic particles.
The lack of a sharp distinction between the Higgs and confined phases in SU(2) theories
with a scalar in the fundamental has been proved by
Fradkin, Shenker et al.\ ~\cite{Osterwalder:1977pc,Fradkin:1978dv,Banks:1979fi,0901.4429}.
A detailed analysis of how this surprising duality applies to the Standard Model
can be found in~\cite{Abbott:1981re,hep-th/9812204}
(we now know that the SU(2)$_L$ gauge group is weakly coupled,
so that in the SM this duality has no physical interest).

We will find extra examples of Higgs/confinement dualities, 
and propose a general criterion:
such a duality holds when a scalar $\S$
in a representation $R$ can break the gauge group $\G$  
to a unique sub-group $\H$ (and thereby with a Higgs phase that is unique). 
In these cases $\S$ admits a single quartic coupling, and
the broken theory contains a single Higgs scalar, that we call $s$.
This happens when $\S$ fills a fundamental of the
$\SU(\ndc)$, $\SO(\ndc)$, $\Sp(\ndc)$, $G_2$ groups
(up to equivalences).
While in the original model \cite{0811.0172,0907.1007,1306.2329} $\G=\SU(2)$ gets fully broken,
in our examples $\H$ has a non-trivial gauge dynamics -- its own confinement -- that
must be taken into account.
On the other hand, a scalar in the fundamental of $F_4$, $E_6$, $E_7$, $E_8$ , or in a higher representation of any group, such as a spinorial of SO(10), 
instead has multiple quartic couplings and gives inequivalent breaking patterns, 
leaving extra scalars in the broken theory.

We will here study theories that satisfy the Higgs/confinement duality, 
and their application to DM.
Such theories can be seen as extensions of those previously listed in 1), 2), 3), and
give qualitatively new physics.
We consider one elementary scalar $\S$ in the  fundamental representation
of a gauge group $\G$ with 
vectors $\mathcal{G}^a_{\mu\nu}$ in the adjoint.
We consider the most generic renormalizable Lagrangian\footnote{The dark gauge group
can have an extra topological term.
In the absence of fermions, it cannot be rotated away.
Such term would violate CP at non-perturbative level.
The $\SU(\ndc)$, $\SO(2\ndc)$ and $E_6$ groups
with symmetric Dynkin diagrams 
admit a $\mathbb{Z}_2$ outer automorphism
(complex conjugation)~\cite{1608.05240} that
acts on vectors by flipping the sign of some vectors,
as determined by the vanishing of some $f^{abc}$ group structure constants.
More simply, the CP-even vectors are those associated to purely imaginary generators $T^a$ in 
some complex representation (e.g.\  fundamental or spinorial).}
\be\label{eq:Lag}
\La=\La_{\rm SM}-\frac{1}{4} {\G}^a_{\mu\nu}{\G}^{a\,\mu\nu} -V_\S+
\begin{cases}
|\D_\mu\S|^2 & \hbox{if $\S$ is complex,}\\
(\D_\mu\S)^2/2 & \hbox{if $\S$ is real,}\end{cases}
\ee
with scalar potential
\beq  V_\S =\begin{cases}
-M_\S^2 |\S|^2 +  \lambda_\S |\S|^4 -  \lambda_{H\S} |H|^2 |\S|^2 
& \hbox{if $\S$ is complex,}\\
- M_\S^2 \S^2/2 +  \lambda_\S  \S^4/4 -  \lambda_{H\S} |H|^2 \S^2/2  & \hbox{if $\S$ is real.}\end{cases}
\eeq
$\S$ is complex when $\G=\SU(\ndc)$ or $\Sp(\ndc)$: in such cases the
the theory is invariant under an accidental U(1) global symmetry, dark baryon number, that rotates the phase of $\S$.
$\S$ is real when $\G=\SO(\ndc)$ or $G_2$: we will discuss the accidental symmetries
of these theories.
These minimal theories give non-trivial DM physics.

If $\G$ confines, baryons made of scalars $\S$ are stable DM candidates.
As we will see, their nature qualitatively depends on the group $\G$. If $\S$ gets a vacuum expectation value, $\G$ gets broken
to a subgroup $\H$,
\beq \SU(\ndc)\to \SU(\ndc-1),\quad
\SO(\ndc)\to \SO(\ndc-1),\quad
\Sp(\ndc)\to\Sp(\ndc-2),\quad
G_2\to \SU(3),\eeq
and some massive vectors are accidentally stable DM candidates.
At lower energy $\H$ confines,
giving rise to various states (dark glue-balls, dark mesons, ...) 
and to baryonic DM, in such a way that
the Higgs/confined and $\G$-confined phases are equivalent.\footnote{In order to avoid confinement,
\cite{1505.07480,1611.00365} considered non-minimal 
models with enough multiple scalars that $\SU(\ndc)$ gets broken to nothing.
We accept condensation and focus on the minimal scalar content.
Given that DM is the lightest stable particle, this also approximates
the DM phenomenology of more general theories provided that the
extra particles are heavier at least by $\Delta M \circa{>}\LDC$,
where $\LDC$ is the scale at which $\gDC $ becomes strongly coupled, if unbroken.}



Models with a new confining gauge group $\G$ and new fermions ${\cal F}$ have been explored in~\cite{1503.08749,1707.05380,1811.06975}: in such models communication
with the SM arises if ${\cal F}$ is charged also under the SM gauge group:
models need to be selected such that the composite DM is neutral.
In scalar models, instead, we can assume that $\S$ is neutral under the SM
(resulting into a neutral DM candidate) because $\S$
interacts with the Higgs through the mixed scalar quartic $\lambda_{H\S}$.



\smallskip

The paper is structured as follows.
In section~\ref{sec:SU} we consider the group $\G=\SU(\NDC)$ and study both the Higgs and condensed phases, focusing on their equivalence, on the accidental symmetry that protects the stability of DM, as seen by both the dual phases, and on DM phenomenology. We then extend the analysis to the other groups for which we find that the duality holds: $\SO(\NDC)$ (section~\ref{sec:SO}), $\Sp(\NDC)$ (section~\ref{sec:Sp}) and $G_2$ (section~\ref{sec:G2}). 
Conclusions are finally given in section~\ref{concl}, where we summarize our main results.

\section{A fundamental of SU($\ndc$)}\label{sec:SU}

\subsection{SU: Higgs phase}\label{Nhiggsed}
Independently of whether symmetry breaking happens dynamically,
in the Higgs phase $\S$ can always be written as
\beq \label{eq:S}
\S(x) = \frac{1}{\sqrt{2}} \begin{pmatrix}
0 \cr \vdots \cr 0\cr w+s(x)
\end{pmatrix}\eeq
such that the gauge group $\SU(\ndc)$ gets broken to $\SU(\ndc-1)$,
leaving one degree of freedom $s$ in $\S$.
While $\med{\S}$ breaks dark baryon number $\U(1)_{\rm DB}$
(under which $\S$ has charge $1$),
a stable DM candidate remains thanks to an
accidental global U(1) symmetry.
Its generator 
$\ndc ( 1, \ldots, 1,0)/(\ndc-1)$
is the unbroken 
linear combination of U(1)$_{\rm DB}$
and the broken U(1) gauge symmetry in $\SU(\ndc)$ corresponding to
the generator (cf.~Appendix \ref{app:gen})
\beq \label{eq:Tdiag}
T^{\ndc^2-1} = \diag (1,\ldots,1 ,1-\ndc)/\sqrt{2\ndc(\ndc-1)}.\eeq
Here and in the following, we normalize $\SU(\ndc)$ generators in the fundamental representation as $\Tr(T^a T^b) = \frac{1}{2} \delta^{ab}$.
It is especially interesting to
consider dynamical symmetry breaking through the Coleman-Weinberg mechanism, obtained by setting $M_\S=0$.
Assuming that $\lambda_{H\S}$ is negligibly small,
the scalar $\S$ dynamically acquires a vacuum expectation value $w = s_* e^{-1/4}$
where $s_*$ is the Renormalization Group Equation (RGE) scale $\mu$ at which 
the running quartic coupling $\lambda_\S(\mu)$
crosses 0, becoming negative at low energy, in view of its RGE at one loop
\begin{eqnsystem}{sys:RGEsym}
(4\pi)^2 \frac{d\gDC}{d\ln\mu} & = & -\frac{22\ndc-1}{6} \gDC^3\,,\\
(4\pi)^2 \frac{d\lambda_\S}{d\ln\mu} &=&\frac34 (\ndc-1) \left(1+\frac{2}{\ndc}-\frac{2}{\ndc^2}\right)\gDC^4 -
6\ndc \gDC^2 \lambda _\S \left(1-\frac{1}{\ndc^2}\right)+4(4+\ndc)  \lambda^{ 2} _\S\, .~~~
\end{eqnsystem}
In such a case the scalar $s$ is known as `scalon' \cite{Gildener:1976ih}  and its 
mass squared is one-loop suppressed, $M_s^2 = w^2 \beta_{\lambda_S}$,
with $\beta_{\lambda_S}\equiv d\lambda_S/d\ln\mu$~\cite{1306.2329}.
If the Higgs mass term is absent too, this model can also generate the 
weak scale $v$, where $v\approx 246\GeV$ is the needed Higgs vacuum expectation value.
Assuming a small positive $\lambda_{HS}$, the weak scale is generated 
as $v \approx w \sqrt{\sfrac{\lambda_{HS}}{2\lambda_H}}$~\cite{1306.2329}.
More complicated expressions hold if $\lambda_{HS}$ is not negligibly small.

The renormalizable action is also invariant under a separate charge conjugation in the dark sector,
that acts as $\S\to\S^{\dagger}$ and flips the sign of $\SU(\ndc)$ vectors with purely real
generators, leaving invariant the vectors with purely imaginary generators.


\medskip

Writing the gauge bosons as 
\beq 
T^a \G^a_\mu = 
\left(
\begin{array}{c|c}
 \A_\mu  &  \W_\mu/\sqrt{2}  \\ \hline
 \W^{*}_\mu/\sqrt{2} & 0 \\
\end{array}
\right) \,-\, \Z_\mu \sqrt{\frac{\NDC-1}{2\NDC}} \left(
\begin{array}{c|c}
 - \One/(\NDC-1)  & 0  \\ \hline
0& 1 \\
\end{array}
\right)
\eeq
the perturbative spectrum is:
\begin{itemize}
\item the C-even scalon $s$, singlet under $\SU(\ndc-1)$, with mass $M_s$;
\item $\ndc(\ndc-2)$ massless dark gluons $\A_\mu$ in the adjoint of $\SU(\ndc-1)$. 
They inherit their $\SU(\ndc)$ transformations under charge conjugation;

 
\item $2(\ndc-1)$ massive dark $\W_\mu$ in the $(\ndc-1)+\overline{(\ndc-1)}$ of $\SU(\ndc-1)$
with mass $M^2_{\W}~=~\gDC^2 w^2/{4}$.
The $\W$ are stable because charged under the global unbroken U(1).
Furthermore they transform as $(i\W)\to(i\W)^*$ under charge conjugation;

\item a massive C-odd dark $\Z_\mu$ corresponding to the generator $T^{\ndc^2-1}$
with mass {$M_\Z^2 =\gDC^2w^2(\ndc-1) /2\ndc$} 
that decays
into $\A\A\A$ at one loop.
\end{itemize}
The case $\ndc=2$ of this model
was studied in~\cite{0811.0172,0907.1007}: the dark gluons do not exist,
and $\W,\Z$ are degenerate thanks to a custodial symmetry.
We consider $\ndc>2$ such that $M_\W < M_\Z < \sqrt{2} M_\W$.
The DM candidate is the $\W$, that undergoes $\W\W^*\to \A\A, ss, s\A,s\Z,\A\Z$ annihilations, while
co-annihilations $\W \Z \to \W s$~\cite{0811.0172} become irrelevant because $\Z$ is not DM for $\ndc>2$.
\subsubsection*{Condensation of $\SU(\ndc-1)$}
The case $\ndc>2$ is qualitatively different from $\ndc=2$
because
the vectors $\A$ confine at a scale $\LDC$
that can be exponentially smaller than $M_\W$:
 \be
\LDC\approx M_\W \exp\left[-\frac{6\pi}{11\,(\ndc-1)\,\adc(M_\W)}\right]
\label{eq:LDC}
\ee
where $\adc(M_\W)$ is the value of the dark gauge coupling at the $\W$ mass.
The squared masses of dark-colored particles receive extra contributions of order $\LDC^2$.
After the condensation of $\SU(\ndc-1)$, the spectrum of the theory contains:
\begin{itemize}
\item C-even dark glue-balls such as $\A\A$ (with mass $\mDG \sim 7 \LDC$~\cite{hep-lat/9901004}) and $f^{abc}\A_{\mu\mu'}^a\A_{\nu\nu'}^b\A_{\rho\rho'}^c$;
\item C-odd glueballs $d^{abc}\A_{\mu\mu'}^a\A_{\nu\nu'}^b\A_{\rho\rho'}^c$, stable thanks to C-parity;
\item the dark scalon $s$;
\item the $\Z$;
\item scalar mesons $\W \W^*$ that decay through
the annihilation of their constituents;
\item   dark baryons $\B \sim \W^{\ndc-1}$, that remain as stable DM thanks to the global $\U(1)$. 
\end{itemize}

For $\LDC \ll M_\W$
the dark baryon spectrum can be computed from non-relativistic quantum mechanics.\footnote{The lightest dark baryon is obtained by minimizing the angular momentum in the spatial part of its wave-function, compatibly with its symmetry under the exchange of two constituents.
For a baryon the dark-color part is totally anti-symmetric, so the product of the spin and spatial wave-functions
must also be totally anti-symmetric. For $\ndc = 3,4$ the spin part alone can be anti-symmetrized, so that the ground state can have a totally symmetric $s$-wave. For higher values of $\ndc$, the spatial wave-function cannot be symmetric, and some orbital angular momentum must be involved.} 
In order to later address the condensed phase $\LDC \sim M_\W$
(where the non-relativistic approximation does not hold)
we  here compute the dark baryon spectrum
using a less usual formalism:
by constructing gauge-singlet operators made of the constituent fields $\W^I_\mu$ and their covariant derivatives that interpolate between the dark baryon and the vacuum. For $\Lambda_{\rm DC} \ll M_{\W}$ we can keep the leading operator in the non-relativistic expansion in the velocity
$\beta \sim p/M \sim \Lambda_{\rm DC}/M_{\W}$, under which the temporal index of vector fields as well as spatial derivatives are suppressed by $\beta$.
Here we discuss the lowest values of $\ndc$ case by case:
\begin{itemize}
\item[$\diamond$] For $\ndc=3$ the lightest baryon is a vector because the
anti-symmetric spin wave-function is obtained as $({ 3}\otimes { 3})_{\rm antisym} = { 3}$. 
Indeed, a combination at leading order in $\beta$ is $\B_{\mu} = \epsilon_{IJ} \epsilon^{\mu \nu \lambda \rho} \W_{\nu}^I \D_\lambda \W_{\rho}^J $, which gives a spin-1 dark baryon. The leading spin-0 and spin-2 operators vanish because of anti-symmetry.

\item[$\diamond$] For $\ndc=4$ the ground state is a scalar because the anti-symmetric spin wave-function is obtained as $(3\otimes 3\otimes 3)_{\rm antisym} = 1$.
The associated dark-baryon operator is, for instance, the scalar 
$\B = \epsilon_{IJK} \epsilon^{\mu\nu\rho\sigma} \W_\mu^I \W_\nu^J (\D_\rho \W_\sigma)^K $. The vector operator $\epsilon_{IJK} \epsilon^{\mu\nu\rho\sigma}\W_\mu^I \W_\nu^J \W_\rho^K $ excites a physical spin-1 resonance only at higher order in $\beta$.
\item[$\diamond$] For $\ndc=5$ the lowest-lying bound state is a spin-1 resonance, corresponding for instance to the operator $\B_\lambda =\epsilon_{IJKL} \epsilon^{\mu\nu\rho\sigma}(\D_\mu \W_\nu)^I \W_\rho^J \W_\sigma^K \W_\lambda^L $. A scalar state arises at higher order in the non-relativistic limit, e.g.\ from the $\mathcal{O}(\beta)$ operator $\epsilon_{IJKL} \epsilon^{\mu\nu\rho\sigma}\W_\mu^I \W_\nu^J \W_\rho^K \W_\sigma^L$. 
\item[$\diamond$] For $\ndc=6$ two derivatives are needed and at leading order the combination $\B_{\lambda} = \epsilon_{IJKLM} \epsilon^{\mu \nu \rho \sigma} \epsilon^{\lambda \tau \eta \xi}  \W_\mu^I \W_\nu^J \W_\rho^K (\D_\sigma \W_\tau^L)  (\D_\eta \W_\xi^M) $ is possible, which gives a spin-1 state.
\end{itemize}

Similar considerations can be done for higher $\ndc$. In each case, different resonances are split by an amount $\Delta M \sim \alpha_{\rm DC}^2 M_{\W}$.
For some values of $\ndc$, bound states with different spin exist at leading order; their fine-structure mass splitting is $\Delta M \sim \alpha_{\rm DC}^4 M_{\W}$.

\subsection{SU: condensed phase}\label{Ncondensed}
In the discussion above, we studied the {\em Higgs} phase of a
theory with
a $\SU(\ndc)$ gauge group and a scalar $\S$ in the fundamental
representation.
The same theory admits an apparently different {\em confined} phase,
where the gauge group $\SU(\ndc)$ 
becomes strong at energies $\LDC \circa{>} M_\S$
(confinement happens before Higgsing)
such that a $\med{\S\S^*}$ condensate forms, rather than a vacuum expectation value.

While the strong dynamics of scalars is mostly unknown,
Fradkin et al.~\cite{Osterwalder:1977pc,Fradkin:1978dv,Banks:1979fi,0901.4429} 
claim a non-trivial theorem:
in the presence of a scalar in the fundamental
there is no sharp distinction between the confined phase and the Higgs phase.
This means that the same asymptotic particles appear in the spectrum ---
a surprising result given that the two phases naively look different
(for example, dark baryon number is unbroken in the confined phase).
More in general, asymptotic states
(even in the Higgs phase) should be described through gauge-invariant operators.
This state/operator
association has practical use in lattice computations
and is interesting from
a formal point of view as a way to describe physics in an explicitly gauge-invariant way,
in particular avoiding splitting fields as a fluctuation over a vacuum expectation value.

The case with $\ndc=2$ has been explicitly
discussed in~\cite{0907.1007},\footnote{The SM provides
a more complicated example of a $\SU(2)$ theory with extra fermions:
the equivalence of the Higgs and confined phases has been discussed in~\cite{Abbott:1981re,hep-th/9812204}.}
showing that both phases lead to a real scalar $s$ and to 3 degenerate
massive vectors $\Z_\mu,\W_\mu,\W_\mu^*$.
As discussed above, 
in the Higgs phase $s$ is the radial part of $\S$, and $\W_\mu,\Z_\mu$ are
the vectors of $\SU(2)_{\rm DC}$.
In the confined phase asymptotic states (bound states)
are associated to gauge-invariant
(singlet) operators.
The same asymptotic states are recovered as~\cite{0811.0172}
\beq
s = \S\S^\dagger,\qquad
\Z_\mu =\S^\dagger\D_\mu \S,\qquad
\W_\mu = \S\D_\mu \S,\qquad
\W_\mu^*=\S^\dagger\D_\mu \S^\dagger,\eeq
with $M_\W =M_\Z$ dictated by a custodial symmetry.
The $\W$ states involve contractions with the SU(2)-invariant $\epsilon_{ij}$ tensor
and thereby
are replaced by $S^\ndc$ baryons
for $\ndc>2$.
Indeed, $\ndc>2$ leads to a  more complicated dynamics
in the Higgs phase, due to the unbroken
$\SU(\ndc-1)$ that confines at lower energy
(the relevance of this confinement
for the validity of the theorem was stressed in~\cite{Dimopoulos:1980hn}).
The main states found in the confined phase match those
found in the Higgs phase as follows:\footnote{In the case of $\SU(3)_{\rm DC}$ 
the validity of the theorem has been tested through
lattice computations~\cite{1607.05860,1709.07477,1804.04453}, with a puzzling result.
Even in a weakly coupled theory the physics in the
gauge-invariant formalism seems different from the physics
found in the standard formalism, where scalars are split
in a gauge dependent way as fluctuations over a vacuum expectation value.
For $\ndc=2$ (for example, in the SM) the difference is claimed
to be in details of cross sections; for
$\ndc=3$ the difference is claimed to be already at the level of the 
spectrum of asymptotic states.
We follow the standard procedure.}
\begin{itemize}

\item C-even and C-odd glue-balls that match those in the Higgs phase;
\item The {C-even} scalar $\S^\dagger\S$ corresponds to $s$;
\item The {C-odd} vector $\S^\dagger \D_\mu \S$ corresponds to $\Z_\mu$;
\item Baryon states $\B\sim \S^\ndc$ can be constructed as follows:
in view of the contraction with the totally anti-symmetric tensor with $\ndc$ indices,
a non-zero contribution is obtained when at least $\ndc-1$ terms contain covariant
derivatives $\D$ acting on the corresponding $\S$;
the baryonic interpolating operators can thus be written as
\beq \B = \S^I  [\epsilon_{IJK\cdots} (\D^{(n)} \S)^J (\D^{(n')} \S)^K \cdots].\label{eq:BSN}\eeq
Under C-parity, baryons transform as $\B\to \B^*$.
%
\end{itemize}
To see that the condensed baryons $\B\sim \S^\ndc$ correspond to the baryons of the Higgs phase $\B\sim\W^{\ndc-1}$
discussed in section~\ref{Nhiggsed},
we notice that when one component of $\S$ acquires a vacuum expectation value,
the term in the square brackets in eq.\eq{BSN} reduces
to the $\W^{\ndc -1}$ baryons formed in the Higgs phase when the residual $\SU(\ndc-1)$ confines, 
after the identification of the $\ndc-1$ Goldstone bosons $\D_\mu \S^J \leftrightarrow \W_\mu^J$. The baryons of the Higgs and confined phases are in a one-to-one correspondence, consistently with Fradkin-Shenker theorem~\cite{Osterwalder:1977pc,Fradkin:1978dv,Banks:1979fi,0901.4429}.

\subsection{SU: phenomenology}\label{Phenomenology}

As the Higgs and confined phases contain the same asymptotic particles, we perform all computations in the Higgs phase
for weak couplings.

The model has 4 extra parameters beyond the SM: $\gDC$, $\lambda_S$, $\lambda_{HS}$, $M_\S^2$.
For simplicity, we will discuss predictions in the dimension-less limit, 
where the model has only 2 extra parameters beyond the SM, given that $M_\S$ vanishes and that
the weak scale $v$ is generated by the dynamical scale $w$, such that $\lambda_{HS}$ is fixed by
\beq v \simeq w \sqrt{\frac{\lambda_{HS}}{2\lambda_H}}.\eeq
The model contains two DM candidates:  the C-odd glueballs with mass $\sim \LDC$ and
the bound state of $\W$ with mass $(\ndc-1)M_\W$.
We can trade $\lambda_\S$ for $(\ndc-1)M_\W$, and determine its value by assuming
that its thermal  relic abundance reproduces the observed cosmological DM abundance.
Indeed, the C-odd dark glueballs never dominate the DM abundance.
Their abundance is negligible if $\Lambda_{\rm DC}\ll w$,
as they annihilate with larger cross section $\sim 1/\LDC^2$ and are lighter.
In the opposite limit, $\Lambda_{\rm DC}\gg w$, the two abundances are comparable, 
as both DM  candidates have masses of order $\LDC$ and annihilation cross sections of order
$1/\Lambda_{\rm DC}^2$. 
However having two candidates does not change our estimates, given that
we do not compute order unity non-perturbative factors.

At the end, only $\gDC$ remains as a free parameter.
The condensed phase is smoothly obtained in the limit where $\gDC\sim 4\pi/\sqrt{\NDC-1}$ becomes non-perturbative.
The phenomenology is similar to the $\ndc=2$ model of~\cite{1306.2329} with an important difference:
the presence of extra light glue-balls. 

\subsubsection{Relic DM abundance}
The thermal relic DM abundance is determined by various events.

First, the usual decoupling of free $\W$ vectors
takes place at the temperature $T \sim M_\W/25$.
The relic abundance of $\W$ vectors is dictated by
$\sigma v_{\rm rel}$,
the tree-level non-relativistic
$s$-wave annihilation cross section
(averaging over initial spin and gauge components, and multiplying by $\kappa=1/2$ for complex DM particles, or $\kappa=1$ for real DM particles).
Using the Feynman rules collected in Appendix \ref{app:SUNFeyn} and
summing over all annihilation processes we obtain the needed cross sections.

We first write those generic cross-sections that arise whenever DM vectors $\W$ with mass $M_\W$
fill a representation  $R$ under the unbroken group $\H$ with massless vectors $\A$:
\begin{eqnsystem}{sys:sigmavrel}
\sigma v_{\rm rel}(\W\W^*\to \A\A) &=&\kappa\frac{19C_R(4C_R-C_{\rm adj})\gDC^4}{288\pi d_R M_\W^2},\\
\sigma v_{\rm rel}(\W\W^*\to \A s) &=&\kappa\frac{2C_R\gDC^2 }{9\pi d_R w^2},\\
\sigma v_{\rm rel}(\W\W^*\to ss) &=&\kappa \frac{11 M_\W^2}{144\pi d_R w^4}.
\end{eqnsystem}
In the above expressions $d_R$ is the dimension of the representation $R$, $C_R$ is the quadratic Casimir of $R$
(defined as $\delta_{ij} C_R =(T^a_RT^a_R)_{ij}$), 
and $C_{\rm adj}$ is the one of the adjoint of  $\H$.
Specializing to the case $\H=\SU(n)$ with
$n=\NDC-1$, we have $C_{\rm adj}=n$, 
$C_{\rm fund}=(n^2-1)/2n$, $d_{\rm fund}=n$, and $\kappa=1/2$ for the complex fundamental, so that\footnote{As a check of our computation, we verified that all cross sections
scale as $1/s$ in the ultra-relativistic limit, often thanks to cancellations
related to the Higgs mechanism.
The $\W\W^*\to \Z \Z$ process is kinematically closed.
For $\ndc=2$ cross sections involving $\A$ vanish, $\Z$ becomes DM forming a degenerate triplet with $\W$,
and the  result in~\cite{1306.2329} is reproduced taking into account the extra cross sections
$2\sigma v_{\rm rel}(\W\W^*\to ss)=\sigma v_{\rm rel}(\Z\Z\to ss)$ and
$2\sigma v_{\rm rel}(\W\W^*\to \Z s)=\sigma v_{\rm rel}(\W\Z\to \W s)=3\gDC^4/128\pi M_\W^2$.}
\begin{eqnsystem}{sys:sigmavrelSU}
\sigma v_{\rm rel}(\W\W^*\to \A\A) &=&\frac{19\gDC^4(n^2-1)(n^2-2)}{1152\pi n^3 M_\W^2},\\
\sigma v_{\rm rel}(\W\W^*\to \A s) &=&\frac{\gDC^4(n^2-1)}{72\pi n^2 M_\W^2},\\
\sigma v_{\rm rel}(\W\W^*\to ss) &=&\frac{11\gDC^4}{4608\pi n M_\W^2},\\
\sigma v_{\rm rel}(\W\W^*\to \Z s) &=&\frac{\gDC^4(n+1)}{72\pi n^2 M_\W^2}
\left(1 - \frac{M_\Z^2}{4M_\W^2}\right)^3,\\
\sigma v_{\rm rel}(\W\W^*\to \Z \A) &=&
\frac{\gDC^4(n-1)(n+1)^2}{288\pi n^3 M_\W^2}
\left(1 - \frac{M_\Z^2}{4M_\W^2}\right)
\left(19+ \frac{M_\Z^2}{M_\W^2}+\frac{M_\Z^4}{4M_\W^4}\right),
\end{eqnsystem}
where the last two extra cross sections involve the $\Z$ vector present in SU models.
These cross sections are  enhanced by order-one Sommerfeld and bound-state effects, that we neglect.
If $\W$ decoupling were the only process, it 
would leave the present abundance
\beq \label{eq:wimp} \frac{\Omega_{\W} h^2}{0.110} \approx \frac{\sigma v_{\rm cosmo} }{\sigma v_{\rm rel}}\qquad\hbox{where}\qquad
\sigma v_{\rm cosmo} \approx  
2.2\times10^{-26}\frac{\cm^3}{{\rm sec}}.\eeq

Second, at the scale of dark confinement the $\W$ forms either mesons $\W\W^*$ that annihilate,
or baryons $\B \sim \W^{\ndc-1}$ that remain as DM.
The mass fraction in baryons is estimated as~\cite{1707.05380} 
\begin{equation}\label{eq:pB}
\wp_\B\approx \frac{1}{1+2^{\ndc-2}/(\ndc-1)} .
\end{equation}

Third, the $\B$ baryons can annihilate with $\B^*$.
The cross section is enhanced by recombination and 
ranges between the squared Bohr radius of the ground state, $\pi R_\B^2$ 
up to $1/\LDC^2$ depending on which levels get occupied
during the cosmological evolution.
If such annihilations happen at temperatures above the Coloumbian binding energy of these states, the cross section gets enhanced by the thermal size of the occupied levels~\cite{1606.00159,1707.05380}.

Fourth, if scalons $s$ and/or glue-balls $\A\A$ produced by DM annihilations
have a long enough life-time, so that
they dominate the energy density of the Universe while decaying into SM particles,
the reheating effects dilutes the DM density.

As physics is complex, it is useful to show estimates that exhibit the dependencies on
parameters.
The thermal relic DM abundance is
estimated as $Y_{\rm DM}\equiv n_{\rm DM}/s \sim 1/(T_{\rm dec} M_{\rm Pl} \sigma_{\rm ann})$,
by demanding $n_{\rm DM} \sigma \sim H$, where $H$ is the Hubble rate:

\begin{itemize}
\item Perturbative annihilations with $\sigma_{\rm ann} \sim\adc^2/M_\W^2$ decouple at $T_{\rm dec} \sim M_\W/25$
leaving $Y_{\rm DM}\sim M_\W/\adc^2 M_{\rm Pl}$.

\item
If all bound states have large cross section $\sigma_{\rm ann} \sim 1/\LDC^2$
at $T_{\rm dec} \sim \LDC$,
this phase leaves $Y_{\rm DM}\sim \LDC/M_{\rm Pl}$.\footnote{Such a geometric
cross section among bound states leads to DM annihilations only 
if constituents de-excite towards the ground state
before that the bound state gets broken by another collision.
If $M_\W\gg \LDC$  large angular momenta are involved, and 
the cross section gets suppressed by a factor roughly estimated as
$\min(1,\adc^{7/4} \LDC^{5/2}/M_\W^{1/2} T^2)$ in~\cite{1801.01135},
which becomes of order unity only at a temperature $T$
mildly below $\LDC$.
A more precise but more model-dependent study of such issues has been performed in~\cite{1811.08418}.\label{foot:uncertain}}

\item However, DM can survive forming heavy baryons with
$\sigma_{\rm ann} \sim \pi R_\B^2 \sim 1/M_\W^2\adc^2$,  so that
$Y_{\rm DM} \sim \adc^2 M_\W^2/\LDC M_{\rm Pl}$. 

\item If the binding energy of heavy baryons
$E_B \sim \adc^2 M_\W$ is smaller than $T \sim \LDC$,
thermally excited baryons acquire a radius $R_\B\sim T/\LDC^2$
such that $\sigma_{\rm ann} \sim 1/\LDC^2$ at $T \sim \LDC$. 

\end{itemize}
Notice that the DM annihilation cross section predicted by cosmology
(and relevant for DM indirect detection)
depends on model parameters proportionally to the DM mass over the DM decoupling temperature: 
in the usual case this ratio is $\approx 25$ giving the usual value of $\sigma v_{\rm cosmo}$;
a larger cross section arises in our case in the Higgs phase, where $T_{\rm dec}\sim \LDC \ll M_\W$.

\begin{figure}[t]
$$\includegraphics[width=0.45\textwidth]{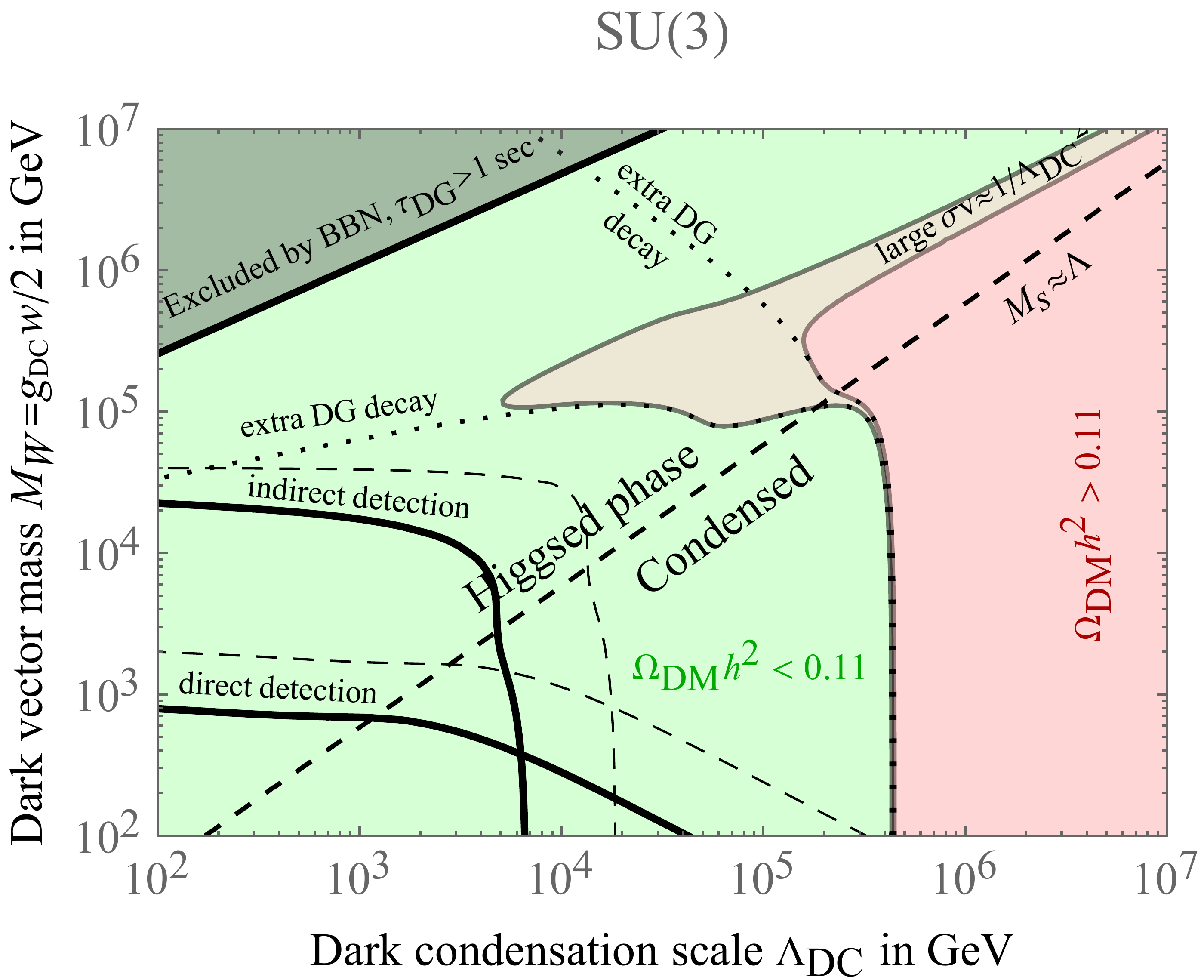}\qquad\includegraphics[width=0.45\textwidth]{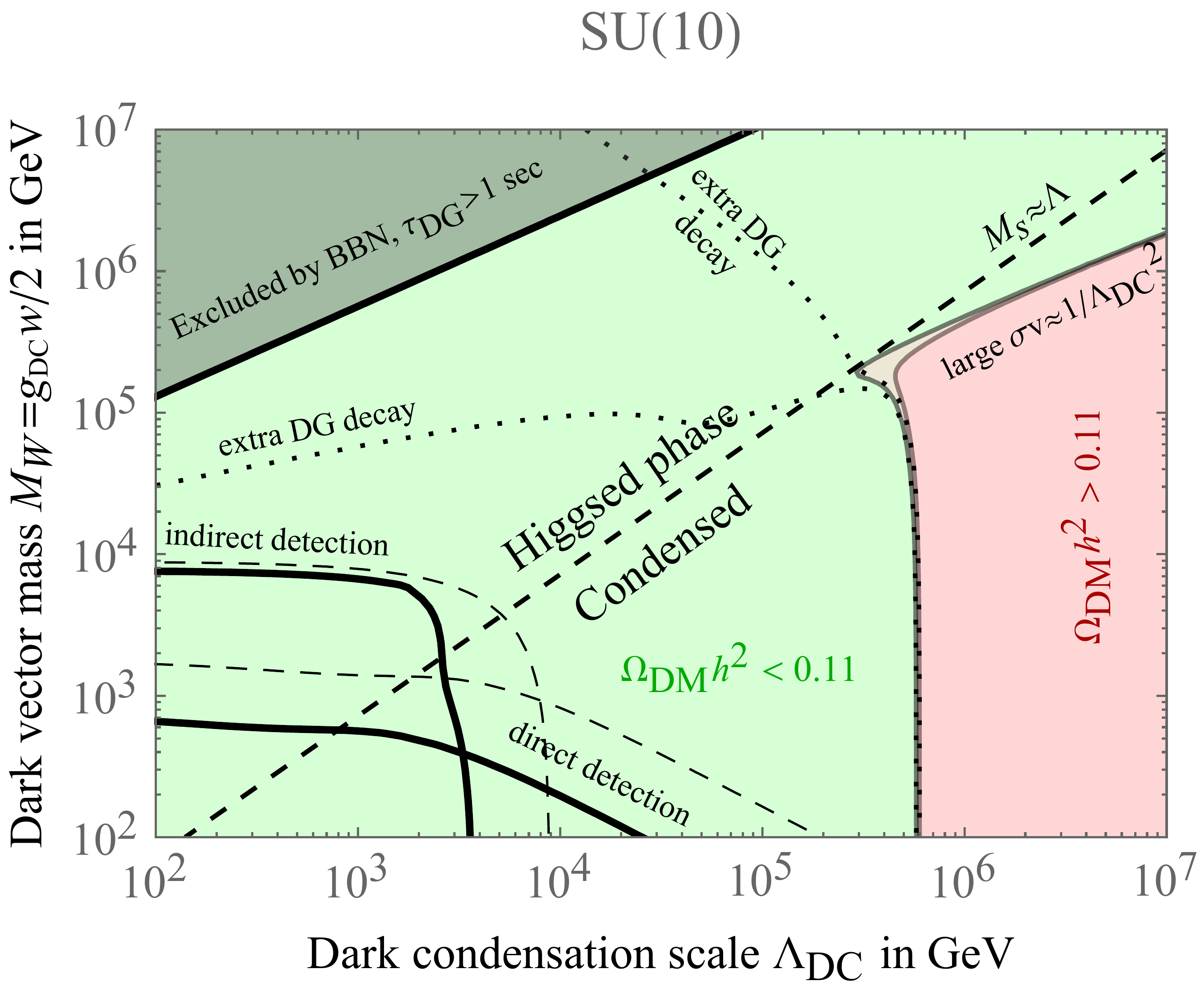}$$
\vspace{-1cm}
\caption{\em \label{fig:DMabundanceSU} 
Physical quantities are here computed as function of the two main scale parameters, 
the dark color confinement scale $\LDC$ and the constituent mass $M_\W =\gDC w/2$.
The boundary between the Higgs and condensed phases
($\gDC\sim 4\pi/\sqrt{n}$ i.e.~$\LDC \sim M_\W$) is shown as dashed line.
The cosmological DM abundance is reproduced along the green/red
boundaries, computed for two different values
of non-perturbative DM annihilation at confinement:
the (thermal) Bohr radius and $1/\LDC^2$.
The latter possibility is relevant only if dark baryons 
do not fall cosmologically fast to their ground state. 
Numerical factors in the condensed phase are fixed matching to
the perturbative cross-sections in the Higgs phase.
We also show, as dotted curves, the  same boundary computed 
assuming that some extra new physics gives fast glue-ball decays. 
The the upper gray region is excluded by BBN because of too slow DG decays and we also show limits from direct~\cite{Aprile:2018dbl} and indirect~\cite{1503.02641} detection (plotted assuming the cosmological DM abundance). {Future prospects are shown as dashed curves}. }
\end{figure}

The predicted relic density is plotted in fig.~\ref{fig:DMabundanceSU} as a function of the two free parameters $\LDC$ and $M_\W$. In the red (green) regions the overall DM abundance turns out to be above (below) the cosmological value, which is reproduced on the boundary between them. We show the results both assuming a Bohr-like annihilation cross-section, as well as a fully non-perturbative $1/\LDC^2$ one. All in all, the observed cosmological abundance $\Omega_{\rm DM} h^2 = 0.11$ can be obtained for DM masses of $\approx 100$ TeV or higher.

\begin{figure}
$$\includegraphics[width=0.45\textwidth]{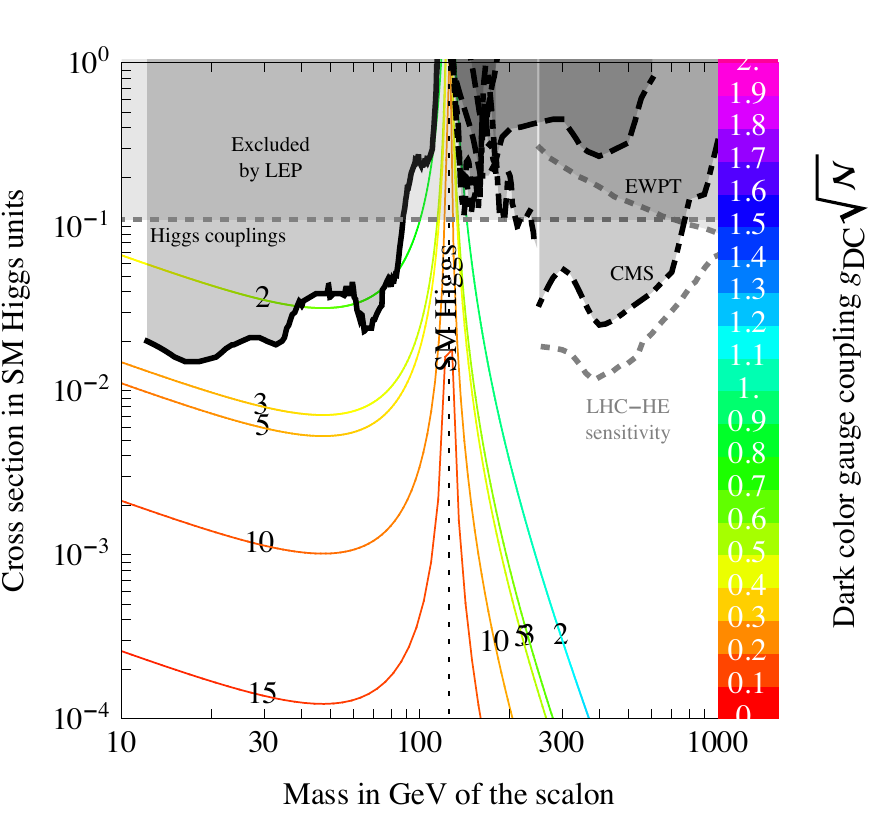}\qquad
\includegraphics[width=0.43\textwidth]{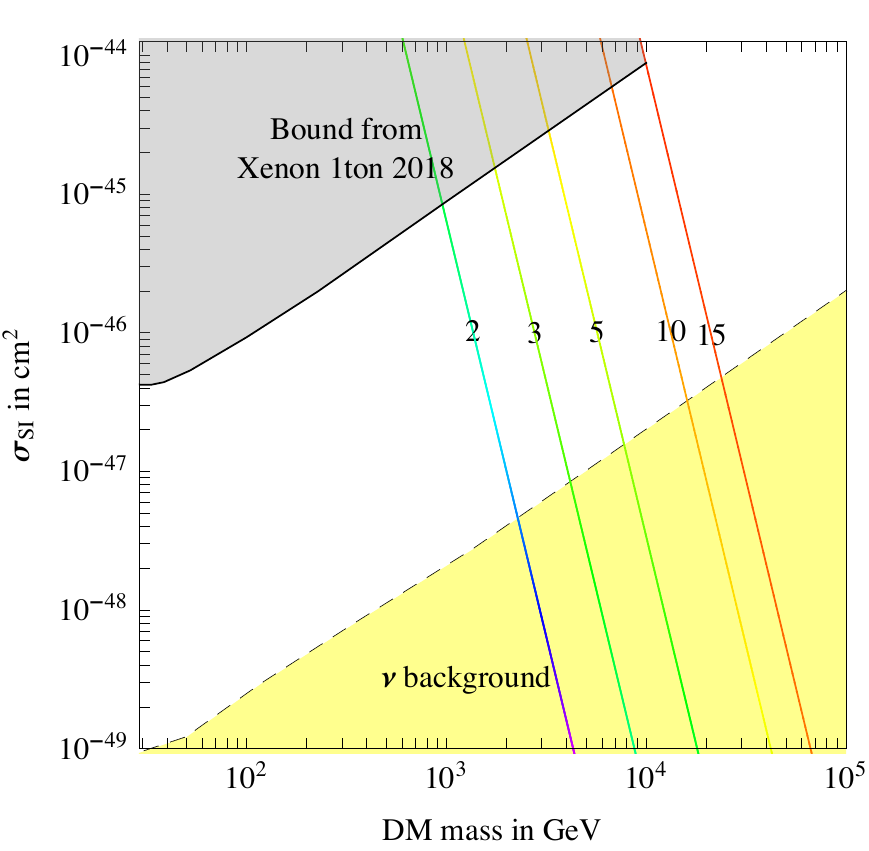}$$
\vspace{-1cm}
\caption{\em \label{fig:phenoN} Predicted cross sections for the extra scalar boson  (left) and for DM direct detection  (right) as function of 
the only free parameter of the model $\gDC$, varied as shown in the color legend. 
We here assume that DM has its cosmological abundance, and that it is
negligibly affected by annihilations among dark baryons. 
The values of $\ndc$ are indicated on the curves. 
}
\end{figure}

\subsubsection{Dark glue-balls}
\label{sec:DG}
{The C-even} Dark Glue-balls (DG) decay through the $\lambda_{HS}$ coupling. We focus on the lightest
$J^{PC}=0^{++}$ glue-balls corresponding to the gauge invariant operator $\Tr[\A_{\mu\nu}\A^{\mu\nu}]$. Heavier C-even glue-balls decay with comparable rates.
The life-time can be computed as follows.
Since the $s$ mass is one-loop suppressed with respect to the $\W$ mass,
 the one-loop effective interaction between $s$ and light vectors $\A$
 can be obtained from their one loop RGE-corrected kinetic term 
\beq 
-\frac{1}{4} (\A^a_{\mu\nu})^2  \left[ 1  - b_\W  \frac{\adc}{4\pi}  \ln \frac{M_\W^2}{\bar\mu^2}\right],\qquad 
\eeq
where $b_\W = -\frac{11}{3}+\frac16 = -\frac72$ 
is the jump in the
RGE  coefficient for $\gDC$ due to $\W$
particles with field-dependent squared mass $M_\W^2 =\frac14 \gDC^2 (w+s)^2$.
Expanding for $s\ll w$  gives the  interaction
\beq \label{eq:soft}
\Lag_{\rm eff} ^{s\A\A}= -  \frac{7\adc}{16\pi }  (\A^a_{\mu\nu})^2 \left(\frac{s}{w} -\frac{s^2}{2w^2}+\cdots\right)\eeq
which results in various decays depending on the mass ordering.
If DG are heavier than the weak scale, they decay into Higgs components as
\beq \Gamma({\rm DG}\to s\to HH^\dagger=hh+ZZ+WW)=\frac{49 f_{\rm DG}^2 \adc^2 \lambda_{HS}^2}{2048 \pi^3 M_{\rm DG} M_s^4}
\Re\sqrt{1- \frac{4M_{h,W,Z}^2}{M_{\rm DG}^2}}\eeq
where $f_{\rm DG}=\med{0|(\mathcal{A}^a_{\mu\nu})^2|{{\rm DG}}} \sim 3 \mDG^3/\gDC^2 \sim \mDG^3$
is a dark matrix element.
The DG life-time can instead become cosmologically large if DG are enough lighter than the weak scale.
If the scalon is heavy, integrating out $s$ gives the  Higgs coupling to dark gluons
\beq
 \Lag_{\rm eff} ^{H\A\A}=  -
 \frac{7\adc \lambda_{HS} }{16\pi M_s^2} (H^\dagger H) ({\cal A}_{\mu\nu}^a)^2,
\eeq
resulting in a mixing angle $\epsilon \ll 1$
between the Higgs and the DG,
and in the consequent DG decay into SM particles 
\beq 
\label{epsGB}
\epsilon \approx \frac{7\adc \lambda_{H\S}vf_{\rm DG}}
{16\pi M_s^2(M_h^2 - \mDG^2)}\, , \qquad
\Gamma_{{\rm DG}} \approx \epsilon^2 \Gamma_{h_{\rm DG}} \, ,
\eeq
where  $ \Gamma_{h_{\rm DG}} $
is the 
decay width 
of a SM Higgs with mass $\mDG$.
More in general, if $s$ can be as light as $h$,
the above expression for $\epsilon$ gets replaced by
\begin{equation}
\epsilon=
\frac{7\adc f_{\rm DG}}{32\pi w} \sin2\gamma
\left(\frac{1}{M_{S_1}^2-M_{\rm DG}^2}-\frac{1}{M_{S_2}^2-M_{\rm DG}^2}\right)
\end{equation}	
where $\gamma$ is the mixing angle that rotates the scalars
$\{h,s\}$ to the mass-eigenstates $\{S_{1},S_{2}\}$ with masses $M_{S_{1,2}}$.
In terms of $v\approx 246\GeV$ it is given by \cite{1306.2329} 
\begin{equation}
\sin 2\gamma = \frac{v^2 \sqrt{8\lambda_H\lambda_{HS}}}{M^2_{S_2} - M^2_{S_1}} \, , 
\end{equation}
with $M^2_{S_2} \approx 2 v^2 \beta_{\lambda_S} \lambda_H / \lambda_{HS}$ 
and $M^2_{S_1} \approx 2 (\lambda_H - \lambda_H^2/\beta_{\lambda_S}) v^2$, 
and where we fix $\lambda_H$ and $\lambda_{HS}$ in terms of the masses from eq.~\eqref{sys:RGEsym}. 
In the limit $M_{S_2}\approx M_s \gg M_h \approx M_{S_1} \gg \mDG$ the width reduces to eq.~(\ref{epsGB}). 
The mixing angle turns out to be small under the assumptions of fig.~\ref{fig:DMabundanceSU} for the relic density. In particular, the region where DM is a thermal relic gives signals much below present bounds.  However the non-perturbative annihilation cross section among DM baryons
 might be smaller than what we assumed
(see the discussion in footnote \ref{foot:uncertain}): sizeable signals
arise in the opposite limit  where annihilations among dark baryons negligibly suppress the
DM abundance.
In fig.\fig{phenoN} we assume that the cosmological DM abundance is set 
by perturbative freeze-out and by the first recombination as in eq.\eq{pB}.
The left plot of fig.\fig{phenoN} shows the predicted value of $\sin^2\gamma$
in this case, which is equal to the production cross-section for the scalon $s$ in SM Higgs cross-section units
(vertical axis), and of $M_s$ (horizontal axis) as function of $\gDC$ (colored legend).  Various present and projected constraints from Higgs measurements and direct searches are also shown \cite{1505.05488,1807.04743}. In particular, it can be seen that measuring Higgs couplings with a $10^{-3}$ precision, which can be attained at future lepton colliders, would allow to probe models where $s$ is light for several values of $\ndc$. 
Similar behaviours are found for the models discussed in the next sections.

\subsubsection{DM indirect detection}
The cross section for $\B \B^*$ annihilations, that gives indirect detection signals, is enhanced by recombination. For small relative velocities $v_{\rm rel}$, it can be written as
\begin{equation}
\sigma_{\B \B^*}v_{\rm rel} \approx \frac{\pi R_\B^2}{\sqrt{M_\B/2E_\B}},
\label{eq:indirectxsec}
\end{equation}
where $R_\B$ is the size of the baryon, $M_\B$ its mass, and $E_\B$ its binding energy. In the limit $\LDC \ll M_\W$  where the constituents are non-relativistic one has
\beq
R_\B \approx \frac{1}{C_{\rm fund} \adc M_\W}, \qquad
 E_\B \approx n C_{\rm fund}^2 \adc^2 M_\W,
\eeq
where we neglected factors of order one.  Inserting the baryon mass $M_\B \approx n M_\W$ we get
\begin{equation}
\sigma_{\B\B^*}v_{\rm rel} \approx \frac{\pi}{C_{\rm fund} \adc M_\W^2}.
\end{equation} 
For larger $\LDC \circa{>} M_\W$ 
the constituents are no longer non-relativistic, and the cross-section becomes $\sigma_{\B\B^*} \approx 1/\LDC^2$.
Annihilations produce scalons and dark glue-balls, that decay into SM particles.

The C-odd dark glue-balls annihilate with cross section $\sigma \sim1/\Lambda_{\rm DC}^2$. 
Their contribution to indirect detection signals is comparable to that of dark baryons if  $\LDC \gg M_\W$
(but in this regime we can only perform estimates)
and negligible if  $\LDC \ll M_\W$ 
(as dark glue-balls annihilate with a larger $\sigma$, 
but the signal rate is proportional to $n_{\rm DM}^2\sigma$
and the glue-ball relic abundance is proportional to $1/\sigma$).
We can thereby ignore C-odd dark glue-balls.

We compare the annihilation cross-section with the 
{HESS limits on a gamma-ray signal in the galactic center \cite{1607.08142}.}
The resulting bound, under the assumption that the dark baryons reproduce the full DM abundance of the Universe, is shown in fig.~\ref{fig:DMabundanceSU} as a function of $\LDC$ and $M_\W$. 
We also show future prospects at CTA \cite{1709.01483}.  However, in the region excluded by indirect detection
the predicted DM abundance is much smaller than the cosmological abundance.
The region where the  DM abundance is reproduced thermally is allowed
by bounds on indirect DM detection.

\subsubsection{Direct detection}
The DM dark baryon $\B$ couples to $s$ (and thereby to $h$)
proportionally to its mass $M_\B \approx (\ndc-1)M_\W$: the effective interaction is
$2 M_\B^2  \,s\B^*\B/w$.
The resulting spin-independent cross section for direct detection is (both for $\B$ of spin 0 and spin 1) 
\begin{eqnarray}
\label{eq:sigmaSISUN}
\sigma_{\rm SI}&=& (\ndc-1)^2\gDC ^2
\sin^22\gamma \frac{m_N^4 f^2}{16\pi v^2} \left(\frac{1}{M_{S_1}^2}-\frac{1}{M_{S_2}^2}\right)^2\\
&\approx&
3.5\, 10^{-44}\cm^2 \times (\ndc-1)^2\gDC ^2 \sin^22\gamma \left(1 - \frac{M_h^2}{M_{S_2}^2}\right)^2,
\end{eqnarray}
where $f\approx0.3$ is a nucleon matrix element.  
In the limit $M_{S_2}\approx M_s \gg M_h = M_{S_1}$ the cross section reduces to
\beq 
\sigma_{\rm SI}\approx(\ndc-1)^2\frac{m_N^4 f^2 v^2 \lambda_H\lambda_{HS}}{2\pi  M_h^4 M_s^4} \gDC ^2 \, .
\eeq
The contribution of the C-odd glueballs to direct detection is negligible if $\LDC \ll M_\W$ 
and comparable to the dark baryon contribution if $\LDC \gg M_\W$ 
(its value is similar to what will computed for an analogous state in the next section).

Fig.\fig{DMabundanceSU} summarizes the situation as function of two
parameters, $\LDC$ and $M_\W$, 
without imposing that the cosmological DM abundance is reproduced. 
We show constraints from Xenon1T \cite{Aprile:2018dbl} as well as
the neutrino floor, that limits  future prospects.
We again see that an allowed region exists, where all signals are significantly below present bounds. In particular the curve where the observed DM abundance is reproduced
lies in the region allowed by all constraints. Here and below, 
we estimated that non-perturbativity 
arises for $g_{\rm DC} \approx 4 \pi/\sqrt{C_{\rm adj}}$,
equal to $4 \pi/\sqrt{n}$ for $\SU(\ndc)$ models.
At such value the Higgs phase smoothly becomes the confined phase.
The DM mass that reproduces the cosmological density in the confined
phase depends on uncertain strong dynamics.
The right plot of fig.\fig{phenoN} shows the predicted value of $\sigma_{\rm SI}$
(vertical axis) and of $M_\B$ (horizontal axis) as function of $\gDC$ (colored legend) 
under the assumption that  DM has its cosmological abundance and that
it is negligibly suppressed by annihilations among DM baryons.

\section{A fundamental of SO($\ndc$)}\label{sec:SO}
The renormalizable Lagrangian is given in eq.\eq{Lag}, with $\S$ real
(since a fundamental of $\SO(\ndc)$ is a real representation).
Thereby there is no accidentally conserved U(1) baryon number.
We normalize the $\SO(\ndc)$ generators in a non-standard way as
$\Tr(T^a T^b) = 2 \delta^{ab}$ in the fundamental
and $\Tr(T^a T^b) = (2\ndc-4) \delta^{ab}$ in the adjoint, in order to keep $\SU(2) \sim \SO(3)$ manifest.
Since the $\SO(\ndc)$ adjoint is the two-index anti-symmetric representation,
the $\SO(\ndc)$ gauge vectors $\G^a$ can be written as $\G_{ij} =\G^a T^a_{ij} $, anti-symmetric under
$ i\leftrightarrow j$.
The RGE are
\begin{eqnsystem}{sys:RGESOn}
(4\pi)^2 \frac{d\gDC}{d\ln\mu} & = & -\frac{22\ndc-45}{3} \gDC^3\,,\\
(4\pi)^2 \frac{d\lambda_\S}{d\ln\mu} &=&6 (\ndc-1) \gDC^4 -
12(\ndc-1) \gDC^2 \lambda _\S +2(8+\ndc)  \lambda^{ 2} _\S\,.
\end{eqnsystem}
\subsection{SO: Higgs phase}
The most generic vacuum expectation value of $\S$
can be rotated to its $\ndc$-th component 
\beq \S(x) =  \begin{pmatrix}
0 \cr \vdots \cr 0\cr w+s(x)
\end{pmatrix}\eeq
such that  SO($\ndc$) is broken to SO($\ndc-1$). 
Writing the gauge bosons as 
\beq 
T^a \G^{a}_\mu = 
\left(
\begin{array}{c|c}
 \A_\mu & -i\W_\mu  \\ \hline
i\W^T_\mu & 0 \\
\end{array}
\right)
\eeq
the perturbative spectrum is:
\begin{itemize}
\item the singlet scalon $s$ with squared mass $M_s^2 = w^2 \beta_{\lambda_S}$; 
\item $(\ndc-1)(\ndc-2)/2$ massless dark gluons $\A_\mu$ in the adjoint of $\SO(\ndc-1)$;
\item $\ndc-1$ real $\W$ vectors in the fundamental of SO($\ndc-1$) with mass $M_\W=\gDC w$.
\end{itemize}
The vectors $\W$ are stable because the
action is invariant under the $\diag(-1,\ldots,-1,1)$ O($\ndc$) reflection
that leaves the $\ndc$-th component of $\S$ invariant,
flipping all other components.
In the broken theory this symmetry acts as
\beq s\to s,\qquad
\W_\mu \to - \W_\mu,\qquad \A_\mu\to \A_\mu.
\eeq
For SO(2) = U(1) dark gluons  $\A$ are absent and this symmetry reduces to U(1)  charge conjugation.


\subsubsection*{Condensation of $\SO(\ndc-1)$}
When $\SO(\ndc-1)$ confines, DM forms dark mesons $\W_i\W_i$ 
(which annihilate), dark glue-balls $\A\A$ (which decay to SM particles), and various baryons,
defined as states formed contracting one $\epsilon_{i_1\cdots i_{\ndc-1}}$ tensor
with the fields of the theory: the heavy $\W_i^\mu$ and the light $\A_{ij}^{\mu\nu}$.
The possibility of using dark gluons as valence constituents of baryons
makes a qualitative difference with respect to the $\SU(\ndc)$ case.
The lightest baryon is the state that contains the lowest possible number of heavy $\W$ and, as discussed in the next sub-section, it is a stable DM candidate:
\begin{itemize}
\item for $\ndc$ even, baryons contain an odd number of $\W$, 
and the lightest baryon contains one $\W$:
\beq\B \sim \W^{i_1} \A^{i_2 i_3} \cdots \A^{i_{n-1} i_{n}}\epsilon_{i_1\cdots i_{n}}\, ;\eeq
\item for $\ndc$ odd, baryons contain an even number of $\W$, and the lightest baryon contains zero $\W$:
\beq \B \sim \A^{i_1 i_2} \cdots \A^{i_{n-1} i_{n}}\epsilon_{i_1\cdots i_{n}}\,.\eeq
\end{itemize}

\subsection{SO: condensed phase}\label{SOcondensed}
We next consider the phase where $\gDC$ is non-perturbatively large,
such that $\SO(\ndc)$ condenses forming the following
singlets under $\SO(\ndc)$:
\begin{itemize}
\item a meson $\S_i\S_j \delta_{ij}$, which is identified with the scalon $s$ in the Higgs phase;\footnote{Note that the gauge part in the operator $\S^T \D_\mu S$ disappears, because of the antisymmetry of the $\SO(\NDC)$ generators. This corresponds to the absence of its corresponding $\Z$ boson in the Higgs phase.\label{foot:Z}}
\item glue-balls $\G\G$, identified with the $\A\A$
glue-balls in the Higgs phase;
\item baryons formed with one $\epsilon_{i_1\cdots i_{\ndc}}$ tensor.
\end{itemize}
Differently from $\SU(\ndc)$ baryons,
the lightest $\SO(\ndc)$ baryon does not need to be made of $\ndc$ fundamentals as
$\epsilon_{i_1\cdots i_\ndc} \S_{i_1}\cdots\S_{i_\ndc}$.
Rather, gauge bosons $\G_{ij}$ can be used to form $\SO(\ndc)$ baryons.
Two constituents $\S_i \S_j$ can annihilate into one dark gluon $\G_{ij}$.
The lightest baryon presumably contains the minimal number of $\S$,
zero or one:
\beq\begin{aligned}
\B \sim \S_{i_1} \G_{i_2 i_3}\cdots \G_{i_{\ndc-1} i_{\ndc}}\epsilon_{i_1\cdots i_\ndc}\qquad &\text{for $\ndc$ odd,}\\
\B \sim \G_{i_1 i_2}\cdots \G_{i_{\ndc-1} i_{\ndc}}\epsilon_{i_1\cdots i_\ndc}\qquad &\text{for $\ndc$ even, $\B^2 \propto\det\G$}.
\end{aligned}\label{eq:BSO}\eeq
These lightest baryons are in one-to-one correspondence with the ones in the broken phase. 
For $\ndc$ odd, when $\S$ gets a vev along its $\ndc$-th component, the remaining $\G$ constituents are identified with the $\A$'s; for $\ndc$ even, exactly one of the $\G$ constituents is identified with the $\W$ in the broken phase due to the $\epsilon$ tensor, while the others are unbroken generators.

The lightest dark baryon is stable because the theory is accidentally invariant under ${\rm O}(\ndc)$ rotations $R_{ij}$
with determinant $-1$~\cite{Witten:1983tx}.  
Acting on baryons $\B$ that contain the $\epsilon$ tensor,
such rotations give $\B \to (\det R)\, \B$,
since
$R_{i_1j_1} \cdots R_{i_\ndc j_\ndc} \epsilon_{i_1\cdots i_\ndc}
=\epsilon_{j_1\cdots j_\ndc} \det R$.
After dividing by $\SO(\ndc)$ rotations one gets a $\mathbb{Z}_2$ symmetry,
that we dub O-parity.
Dark baryons built with one $\ndc$-index anti-symmetric tensor are odd under this $\mathbb{Z}_2$ 
symmetry.
O-parity is a special unusual symmetry, analogous
to space parity and time inversion, in the sense that it can be written in
equivalent explicit ways only after choosing an arbitrary basis in the field space, 
thus fixing one arbitrary rotation with determinant $-1$. 

For $\ndc$ odd, O-parity is more conveniently realised as
a full reflection $-\One$ i.e.\ O-parity acts as
a usual $\mathbb{Z}_2$ symmetry 
\beq\label{eq:Z2SOodd}
\S_i\stackrel{\rm O}{\longrightarrow} - \S_i,\qquad \G_{ij}\stackrel{\rm O}{\longrightarrow} \G_{ij}\eeq
and the lightest baryon is stable because made of an odd number of $\S$'s.

For $\ndc$ even, O-parity can be conveniently realised as 
a reflection under any direction, 
for example along the first component:
$\eta_{1}=\diag(-1,1,\ldots,1)$.\footnote{One can explicitly verify that  this
symmetry is consistent with the Lie algebra of $\SO(\ndc)$
\be 
[T^{ab}, T^{cd}] = i \delta^{ac} T^{bd} + \text{permutations}
\ee
since: if no indices are 1 all generators appearing are even; if one index is 1 both sides are odd; if two indices are 1 both generators on the LHS are odd and the RHS is even.\label{ft:cons}}
As any vector can be gauge-rotated to be along the 1 direction, this morally is parity.
As $\eta_{1}$ anti-commutes with the generators $T^{1i}$ (rotation along the $1i$ plane)
and commutes with the other generators, 
the Lagrangian is invariant under 
\beq\label{eq:Z2SON} \S_1\stackrel{\rm O}{\longrightarrow} -\S_1, \qquad \S_i\stackrel{\rm O}{\longrightarrow} \S_i,\qquad
\G^{1i}\stackrel{\rm O}{\longrightarrow} - \G^{1i},\qquad \G^{ij}\stackrel{\rm O}{\longrightarrow} \G^{ij},\qquad i,j\neq 1.\eeq
Two baryons can annihilate.
The meson $\S\S$ and the glue-balls $\G\G$ are even under O-parity and can decay into SM particles in view of the $\lambda_{HS}$ coupling.

\medskip

For even $\ndc=2N$
the theory contains an extra accidental symmetry different from O-parity.
Indeed $\SO(2N)$ admits one outer automorphism,
which corresponds to complex conjugation C.
Its action on vector bosons can be computed by looking at  the 
generators in the simplest complex representation, the spinor.
C acts as
\beq \G_{ij}\stackrel{\rm C}{\longrightarrow} (-1)^{i+j} \G_{ij},\qquad  \S_i\stackrel{\rm C}{\longrightarrow} (-1)^{i+1} \S_i .\eeq
The consistency with the Lie algebra can be proved analogously to footnote~\ref{ft:cons}. 
Since $\S_i \S_j$ transforms as $\G_{ij}$ the C symmetry does not give extra stable baryons
(despite the fact that for $\SO(4N+2)$ the 0-baryon is odd under C, whereas for $\SO(4N)$ it is even).\footnote{It might seem surprising that C acts non-trivially on $\S$ (a real representation).
This becomes intuitive for $\SO(2) = {\rm U}(1)$: two real scalars are seen as one complex scalar $\S_1+i \S_2$. For larger $N$, the C symmetry similarly reduces to the usual charge conjugation within the $\SU(N)$ subgroup of $\SO(2N)$.  Indeed, since $\U(N) = \SO(2 N) \cap \Sp(2N)$, the $\SU(N)$ subalgebra of $\SO(2N)$ has the form of eq.~\eqref{eq:genSpN}, with $\sigma_k = \sigma_2$. Since $\One_2$ is diagonal, the first set in eq.~\eqref{eq:genSpN} is even under C of $\SO(2N)$; since $T^{(2)}_{\alpha \beta}$ is imaginary, this is even also under C of $\SU(N)$: $T_{\rm imag} \to T_{\rm imag}, T_{\rm real} \to - T_{\rm real}$. Analogously for the remaining two sets in eq.~~\eqref{eq:genSpN}, since $\sigma_2$ is off-diagonal and $T^{(1)}_{\alpha \beta}$ is real.}

%




\subsection{SO: phenomenology}

\subsubsection{Relic DM abundance}
As in $\SU(\ndc)$ models, the thermal relic DM abundance is determined by various cosmological events.

Specializing eq.~(\ref{sys:sigmavrel}) to the unbroken $\H=\SO(n)$ group,
so that $\kappa=1$, $d_{\rm fund}=n=\NDC-1$, $C_{\rm fund}=n-1$ and $C_{\rm adj}=2(n-2)$ in our normalization, 
we obtain 
the annihilation cross sections relevant for the usual decoupling of free $\W$ vectors
\begin{eqnsystem}{sys:sigmavrelSO}
\sigma v_{\rm rel}(\W\W\to \A\A) &=&
\frac{19\gDC^4(n-1)}{144\pi M_\W^2},\\
\sigma v_{\rm rel}(\W\W\to \A s) &=&
\frac{2\gDC^4(n-1)}{9\pi n M_\W^2},\\
\sigma v_{\rm rel}(\W\W\to ss) &=&
\frac{11\gDC^4}{144\pi n M_\W^2}.
\end{eqnsystem}
When $\SO(\ndc-1)$ confines, we need to distinguish two cases:
\begin{itemize}
\item For even $\ndc$ roughly all $\W$'s end up in 1-baryons.
The final DM abundance is set by $\B\B$ annihilations,
as they have a large cross section, of order
$\sigma_{\rm ann} \sim 1/\LDC^2$, giving
 $Y_{\rm DM}\sim \min(M_\W/\adc^2,  \LDC)/M_{\rm Pl}$ for particles with mass of
order $M_\W + \LDC$.

\item For odd $\ndc$ roughly all $\W$'s end up in 2-baryons that decay to 0-baryons,
so that the $\W$ abundance negligibly contributes to the final DM abundance,
approximated by $Y_{\rm DM}\sim \LDC/M_{\rm Pl}$ for particles with mass of
order $\LDC$. Given that in this case the cross-section $\sigma_{\rm ann}\approx 1/\LDC^2$ cannot be extrapolated from a perturbative calculation, the overall coefficient has to be assumed.

\end{itemize}

\begin{figure}
$$\includegraphics[width=0.45\textwidth]{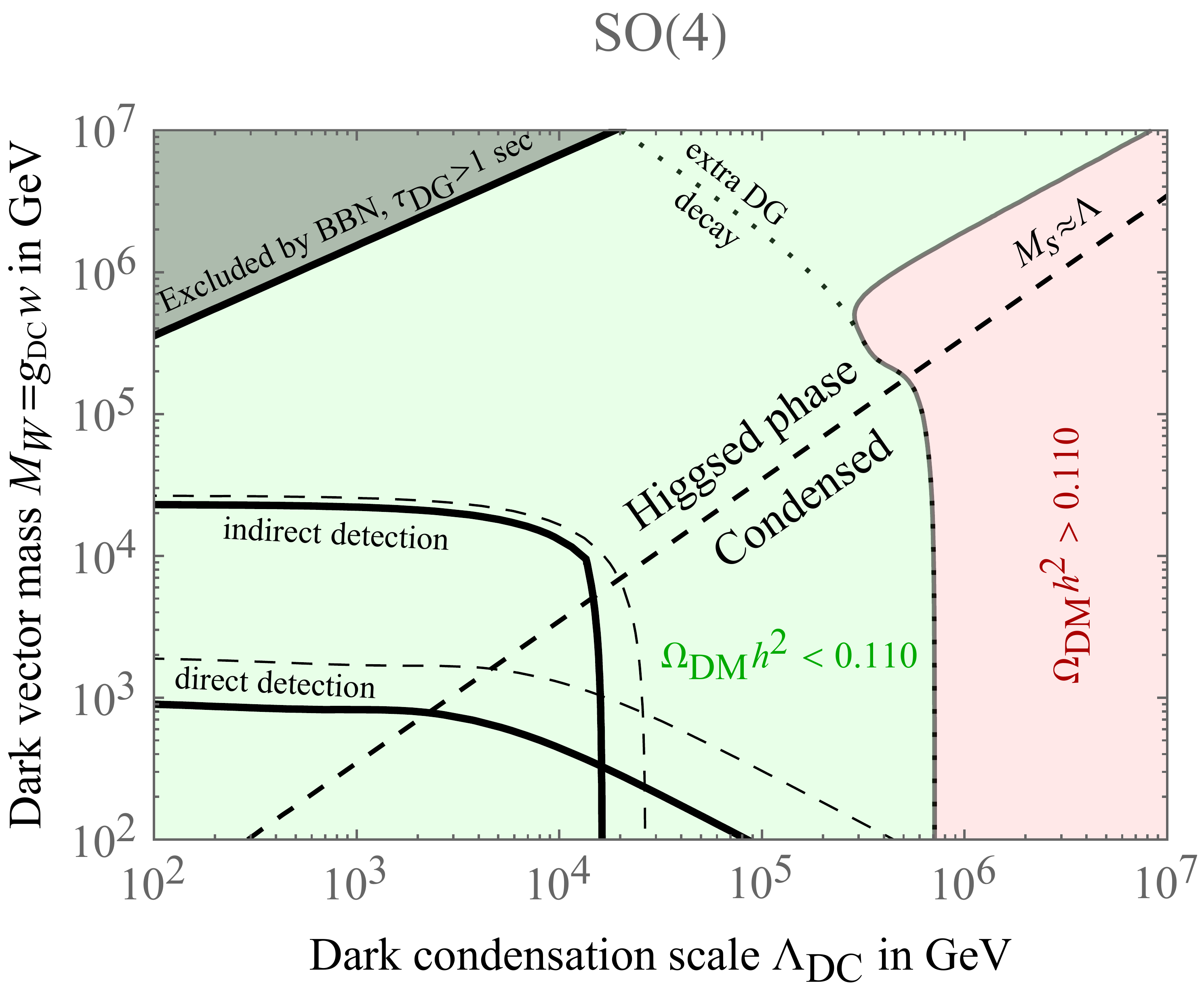} \qquad \includegraphics[width=0.45\textwidth]{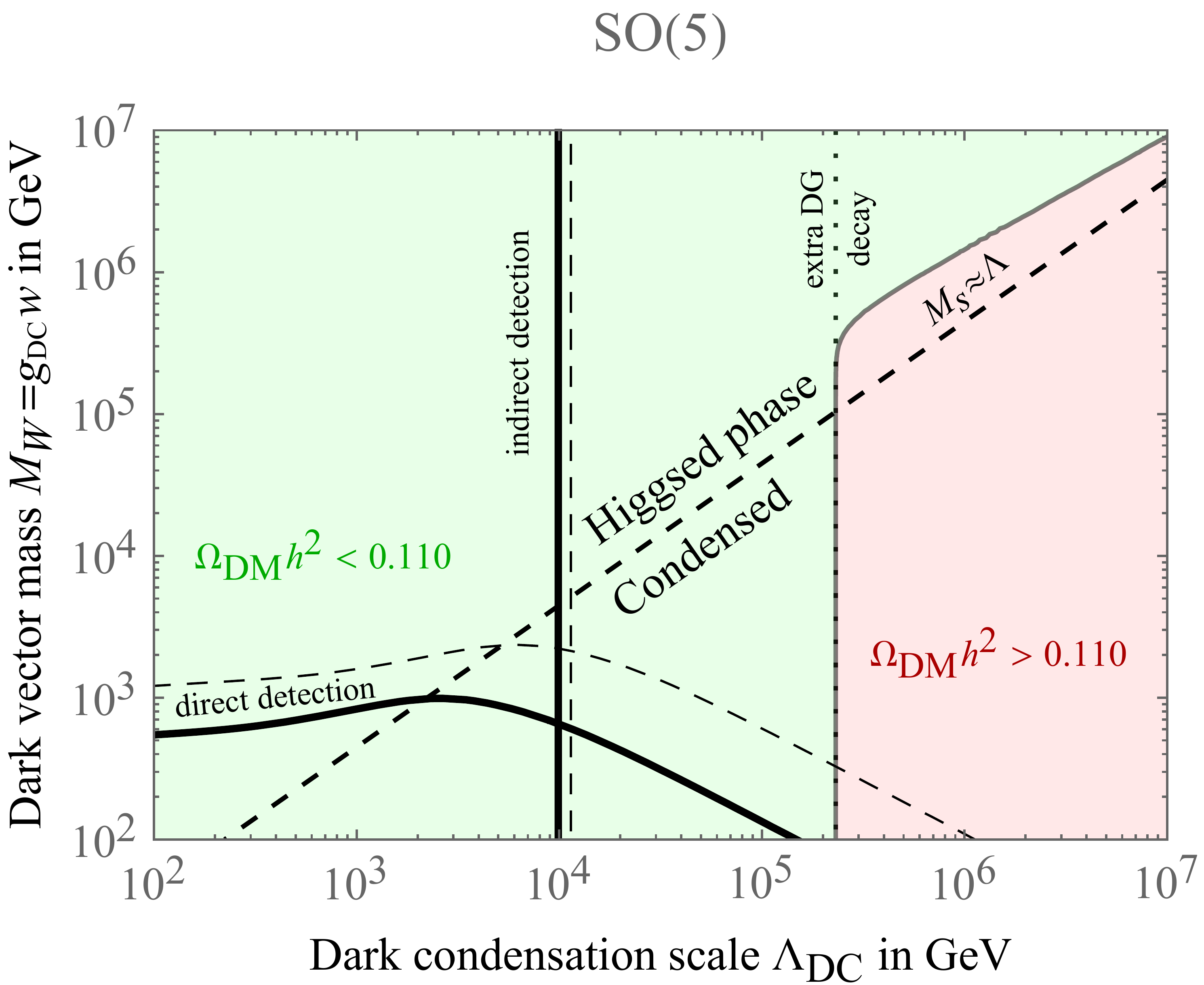}$$
\vspace{-1cm}
\caption{\em \label{fig:DMabundanceSO} 
As in fig.\fig{DMabundanceSU}, for $\SO(\ndc)$ models.
The left plot shows an example with $\NDC$ even
(DM is the baryon with one $\W$ constituent).
The right plot shows an example with $\NDC$ odd
(DM is the baryon with zero $\W$ constituents).
In the right panel, the non-perturbative DM annihilation cross-section cannot be matched to 
its perturbative value: we assume
$\sigma_{\rm ann} \approx 100/\LDC^2$.
}
\end{figure}

The relic abundance is plotted in fig.~\ref{fig:DMabundanceSO} as a function of $\LDC$ and $M_\W$. Notice that, since DM never contains more than one heavy constituent, the relevant annihilation cross-section is always taken to be of order $1/\LDC^2$. Again, the correct DM abundance is obtained for masses of order 100 TeV or heavier.

\subsubsection{DM indirect detection}
The cross section for $\B \B$ annihilations, that gives indirect detection signals, is 
\begin{equation}
\sigma_{\B \B}   \sim  {\cal O}(1)/{\LDC^2},
\label{eq:indirectxsecSO}
\end{equation}
where the ${\cal O}(1)$ factor is $\sim 100$ in the QCD proton case.
DM annihilations produce dark glue-balls, that decay into SM particles.
The HESS bound is plotted in fig.~\ref{fig:DMabundanceSO} as a black line, assuming that the dark baryons constitute the totality of DM. 
Also in this case the region where DM is a thermal relic is not probed.

\subsubsection{DM direct detection}\label{DDSO}
For even $\ndc$, DM contains one heavy
$\W$ constituent 
and the direct detection cross section is
\beq
\sigma_{\text{SI}}=\left(\frac{M_\B}{w}\right)^2\frac{m_N^4f^2}{4\pi v^2}\sin^22\gamma\left(\frac{1}{M_{S_1}^2}-\frac{1}{M_{S_2}^2}\right)^2 \qquad
	\hbox{for $\SO(\ndc)$ with even $\ndc$.}
\eeq 
with $M_\B\simeq M_\W$. 

For odd $\ndc$, DM is a glue-ball that contains zero $\W$ constituents (`odd-ball'), and its direct detection is qualitatively different.
The odd-ball coupling to $s$ --- that mixes with the Higgs --- can be computed
by extending soft theorems.
One-loop RGE running for $E < M_\W < E_0$ can be written as
\beq   \frac{1}{\adc(E)} = \frac{1}{\adc(E_0)} - \frac{b_H}{2\pi} \ln\frac{E}{M_\W}
  - \frac{b_G}{2\pi} \ln\frac{M_\W}{E_0}\eeq 
where $b_H=-22(\ndc-3)/3$ in the broken $\SO(\ndc-1)$ theory,
and $b_G=(45-22\ndc)/3 = b_H + b_\W$ in the unbroken theory, with $b_\W = -7$.
From this we compute how, at fixed high-energy value of the gauge coupling,
$\adc(E_0)$, the scale $E=\LDC$ at which $\adc$ becomes
non-perturbative depends on $M_\W$ and thereby $s$, finding:
\beq
\frac{s}{\LDC} \frac{\partial\LDC}{\partial s} = -\frac{b_\W}{b_H}=-\frac{21}{22(\ndc-3)}= -\frac{7\adc}{2\pi} \ln \frac{M_\W}{\LDC}.\eeq
Then the odd-ball (or glue-ball) mass term $M_\B^2  \B^2/2$ with 
$M_\B\propto \LDC$ gets promoted to\footnote{The factor in parenthesis can be rewritten in terms of $\adc$:
this computation is equivalent to using the usual Higgs soft theorem
of eq.\eq{soft} (that becomes $- (\A^a_{\mu\nu})^2  {7s\adc}/{8\pi w} $ with SO factors)
and using the scale anomaly to compute the baryonic matrix element of such operator.}
\beq \Lag = \frac{M_\B^2}{2} \B^2 \left(1 -\frac{2b_\W}{b_H}\frac{s}{w} + \cdots\right).
\eeq
The resulting direct detection cross section is 
\beq
\sigma_{\text{SI}}=\left(\frac{b_\W}{2b_H}\frac{M_\B}{w}	\right)^2
\frac{m_N^4f^2}{4\pi v^2}\sin^22\gamma\left(\frac{1}{M_{S_1}^2}-\frac{1}{M_{S_2}^2}\right)^2\qquad
\hbox{for $\SO(\ndc)$ with odd $\ndc$.}
\eeq
Fig.\fig{DMabundanceSO} shows  the final results. 
Direct detection could probe a small region where DM is a thermal relic.


\subsubsection{Special cases}
To conclude, we mention some special cases.

For $\ndc=3$ one has $\SO(3)\to \SO(2)$, which is the same
as $\SU(2)\to {\rm U}(1)$ with a scalar in the adjoint.
The U(1) does not confine, leaving long range interactions among DM particles.
Furthermore, topological defects are possible. The viability of such models
will be discussed in a separate publication. 

For $\ndc=4$ the identity $\SO(4) \cong \SU(2)^2$ holds and
the vector $\W$ lies in the adjoint {3} of the unbroken SO(3).

For $\ndc=6$ an extra model with $\S$ in the spinorial of SO(6) breaks univocally to SU(3) leaving the scalon as only scalar; in view of the identity $\SO(6) \cong \SU(4)$ this model has been already discussed as SU(4).

For $\ndc=8$, the group SO(8) has three representations 
with dimension 8 (the fundamental and two spinorials) related by a $S_3$ outer automorphism:
all {8} representations are real and break SO(8) to SO(7).

\section{A fundamental of Sp$(\ndc)$}\label{sec:Sp}
The group Sp$(\ndc)$ is defined for even $\ndc$  
as the transformations that leave
invariant the tensor $\gamma_\ndc\equiv \One_{\ndc/2}\otimes \epsilon$, where
$\epsilon_{ij}$ is the 2-dimensional anti-symmetric tensor.
The fundamental representation of Sp$(\ndc)$ is pseudo-real.
We thereby introduce a complex $\ndc$-dimensional scalar $\S$.
The Lagrangian, given in eq.\eq{Lag}, conserves an accidental baryon U(1),
by virtue of $\S^T \gamma \S=0$.
The adjoint is the trace-less symmetric representation with dimension $\ndc(\ndc+1)/2$.
The RGE are
\begin{eqnsystem}{sys:RGESOn}
(4\pi)^2 \frac{d\gDC}{d\ln\mu} & = & -\frac{11\ndc+21}{6} \gDC^3\,,\\
(4\pi)^2 \frac{d\lambda_\S}{d\ln\mu} &=&\frac{3}{16} (\ndc+4) \gDC^4 -
3(\ndc+1) \gDC^2 \lambda _\S +4(4+\ndc)  \lambda^{ 2} _\S \,.
\end{eqnsystem}
\subsection{Sp: Higgs phase}
Again the RGE gives that $\S$ can radiatively acquire a vacuum expectation value
as in eq.\eq{S}, breaking the gauge group Sp($\ndc$) to Sp($\ndc-2$). 
At the same time, the accidental U(1) global dark-baryon number gets rotated
to an unbroken global U(1) with generator 
\beq \label{eq:TdiagSp}
\diag(1,\ldots,1,1,1) + \diag(0,\ldots,0,1,-1) =  \diag (1,\ldots,1 ,2,0) ,\eeq
obtained combining the original U(1) with the diagonal generator of the broken $\Sp(2)$.
Writing the gauge bosons as 
\beq 
T^a \G^{a}_\mu = 
\left(
\begin{array}{c|cc}
 \A_\mu & \X^{*}_\mu/2  & \gamma_{\ndc-2}\X_\mu/2 \\ \hline
\X_\mu/2 & \Z_\mu/2 &  \W_\mu/\sqrt{2} \\
\gamma_{\ndc-2}\X^*/2 & - \W_\mu^*/\sqrt{2} & -\Z_\mu/2 \\
\end{array}
\right)
\eeq
the perturbative spectrum is:
\begin{itemize}
\item the scalon singlet $s$;
\item one real $\Z$ with mass $M_\Z = \gDC w/2$ and zero dark baryon charge;
\item one complex $\W$, with mass $\M_\W=M_\Z$ and dark baryon charge 2.
\listpart{For $\NDC=2$ this is the $\Sp(2)=\SU(2)$ model of~\cite{0811.0172,0907.1007,1306.2329}
where $\W$ and
$\Z$ are co-stable DM candidates.  For $\ndc\ge 4$ the spectrum contains extra particles:}
\item  $\ndc-2$ complex massive vectors $\X$ in the fundamental representation of $\Sp(\ndc-2)$
with mass $M_{\X}=M_{\W}/\sqrt{2}$ and dark baryon charge 1;
\item the massless vectors $\A$ of $\Sp(\ndc-2)$.
\end{itemize}
The $\Z$ boson decays into $\A$'s.
At perturbative level the $\W$ and $\X$ are DM candidates, co-stable  thanks to accidental baryon  number conservation.
The cubic vector vertices are
\beq \Z\W^*\W,\qquad
\A\A\A,\qquad
\A\X\X^*,\qquad
\X\X\W^*,\qquad
\X\X^*\Z .\eeq


\subsubsection*{Condensation of $\Sp(\ndc-2)$}
When the theory becomes strongly coupled, $\Sp(\ndc-2)$ confines giving the following spectrum of 
asymptotic states:
\begin{itemize}
\item The scalon $s$, the $\Z$ and $\W$ bosons, and dark glue-balls $\A\A$.
\item Two kinds of dark mesons: the unstable $\X^\dagger\X$ and $\X^\dagger \D_{\mu} \X$, which have the same quantum numbers as $s$ and $\Z$, and $\M_\mu=\X^T \gamma_{\ndc-2} \D_\mu\X$,
with dark baryon number 2 as $\W$. Only one linear combination of $\M$ and $\W$ appears among the stable asymptotic states, while the other corresponds to a resonance. 
A similar situation holds for $s$ and $\X^\dag\X$, and for $\Z$ and $\X^\dag\D\X$.

\item Dark baryons $\B$
(defined as states formed with one $\epsilon_{i_1\cdots i_{\ndc-2}}$ tensor)
are not stable because the $\epsilon$ tensor can be decomposed as
$\epsilon_{i_1\cdots i_{\ndc-2}} = \gamma_{i_1 i_2}\cdots \gamma_{i_{\ndc-3}i_{\ndc-2}} + \hbox{permutations}$~\cite{Witten:1983tx}. 
This means that $\B$ splits into $\ndc/2-1$ mesons $\M$.
\end{itemize}
%
Both the $\W$ and the mesons $\M$  carry charge 2 under
conserved U(1) baryon number.


\subsection{Sp: condensed phase}
Confinement of $\Sp(\NDC)$ gives rise to the following bound states:
\begin{itemize}
\item $\S^\dag\S$ corresponding to $s$ and $\X^\dag \X$ (not distinguished by any quantum number);
\item $\S^\dag \D_\mu \S$ corresponding to $\Z_\mu$ and $\X^\dag \D_\mu \X$; 
\item $\S^T \gamma_\ndc \D_\mu \S$ corresponding to $\W_{\mu}$ and $\X^T \gamma_{\ndc-2} \D_\mu \X$;
\item dark glue-balls $\G\G$, corresponding to $\A\A$.
\end{itemize}
The condensed phase coincides with the Higgs phase, in agreement with our generalization of the Fradkin-Shenker theorem. 

\subsection{Sp: phenomenology}

\subsubsection{Relic DM abundance}
At large $\NDC$ the dominant perturbative annihilation cross-section is  
\beq
\sigma v_{\rm rel}(\X\X^*\to \A\A) =\frac{19C_{\rm fund}(4C_{\rm fund}-C_{\rm adj})\gDC^4}{288\pi d_{\rm fund} M_\X^2}
\eeq
with $d_{\rm fund}=2(\ndc-2)$, $C_{\rm fund}=(\ndc-1)/4$, $C_{\rm adj}=\ndc/2$.
Annihilation and semi-annihilation cross sections of the $V=\W,\Z$ vectors into $ss$ and $Vs$
are as in the SU(2) model~\cite{0811.0172}.

As $\X$ annihilate more than $\W$, the latter can have a 
larger relic abundance.
The two sectors ($\W,\Z$ and $\X,\A$) are however coupled by 
$\A \W \leftrightarrow \X\X$ processes, that
thermalize their relative abundances, so that the lighter $\X$
would get a larger abundance than the heavier $\W$.
The final abundance depends on which process decouples earlier:
a detailed computation would be needed.

When $\Sp(\ndc-2)$ condenses, about half $\X$ form stable mesons $\M$.
The cubic vertex $\X\X\W^*$ becomes a $\M \W$ mass mixing.
In the limit  $\LDC\ll w$ their masses are $M_\M =\sqrt{2} M_\W$
so that $\M$ decays to $\W$ and glue-balls (before that 
$\M\M^*$ annihilations with $\sigma_{\rm ann}\sim \pi R_B^2$ deplete the $\M$ abundance),
leaving $\W$ as the DM candidate. 
We estimate that the final DM relic density is approximated by perturbative freeze-out abundance
of $\W$
(up to the suppression present if glue-balls decay slowly when they dominate the energy density).
Hence, differently from the previous cases, in these models dark glue-balls can be so light that they can be probed by collider experiments,
for example by measuring Higgs properties.

\subsubsection{Indirect detection}
The indirect detection cross section is given by the perturbative expressions \eqref{sys:sigmavrelSU} with $n=1$, given that the $\W$ is neutral under the unbroken $\Sp(\ndc-2)$, analogously to the SU(2) model. The final results are shown in fig.~\ref{fig:DMabundanceSpG2} (left panel) for $\ndc = 4$. 
Since the relic density is mainly governed by the perturbative freeze-out of $\W$, the parameter space where DM is a thermal relic can be probed by future experiments.

\subsubsection{Direct detection}
\begin{figure}
$$\includegraphics[width=0.45\textwidth]{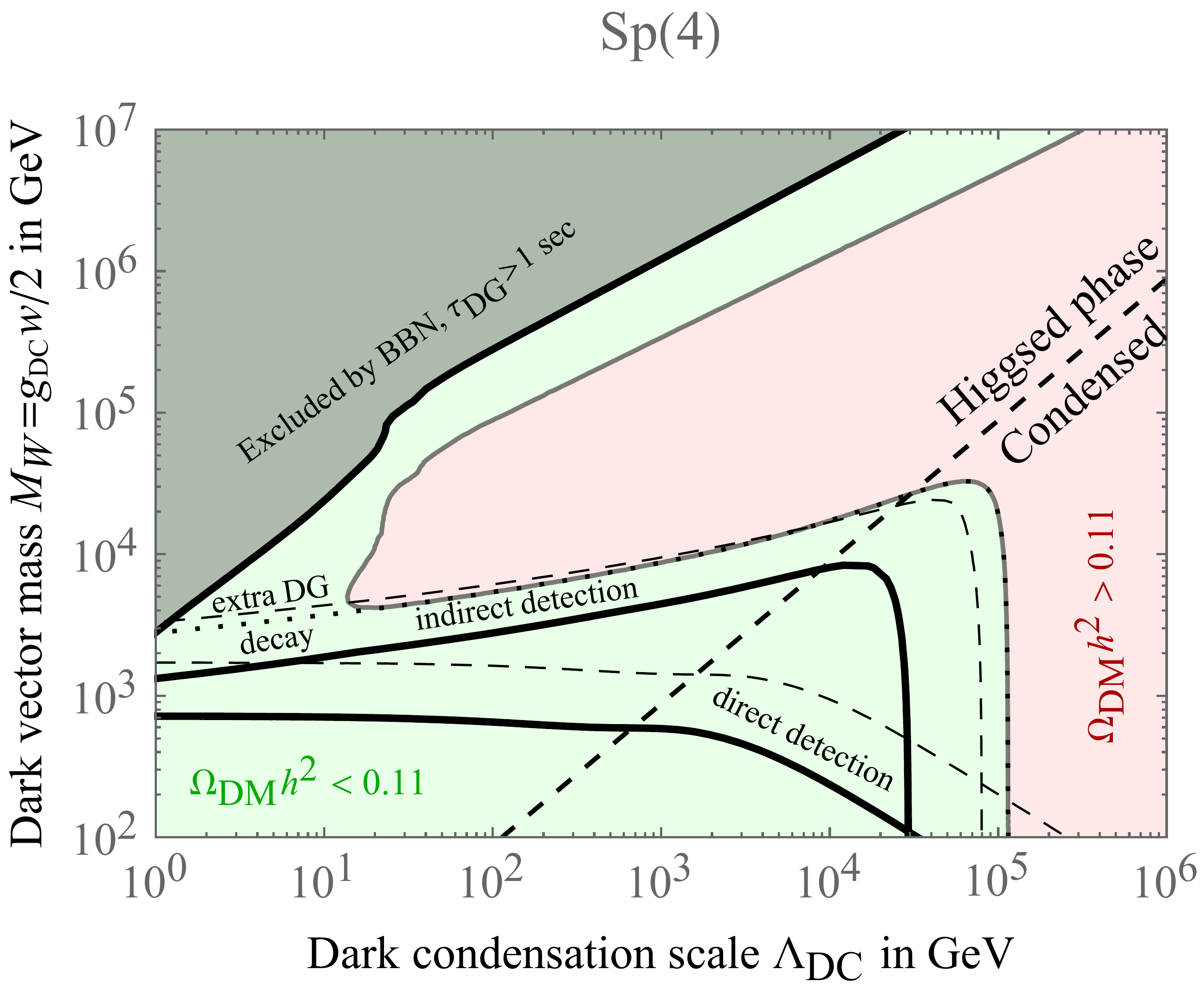} \qquad \includegraphics[width=0.45\textwidth]{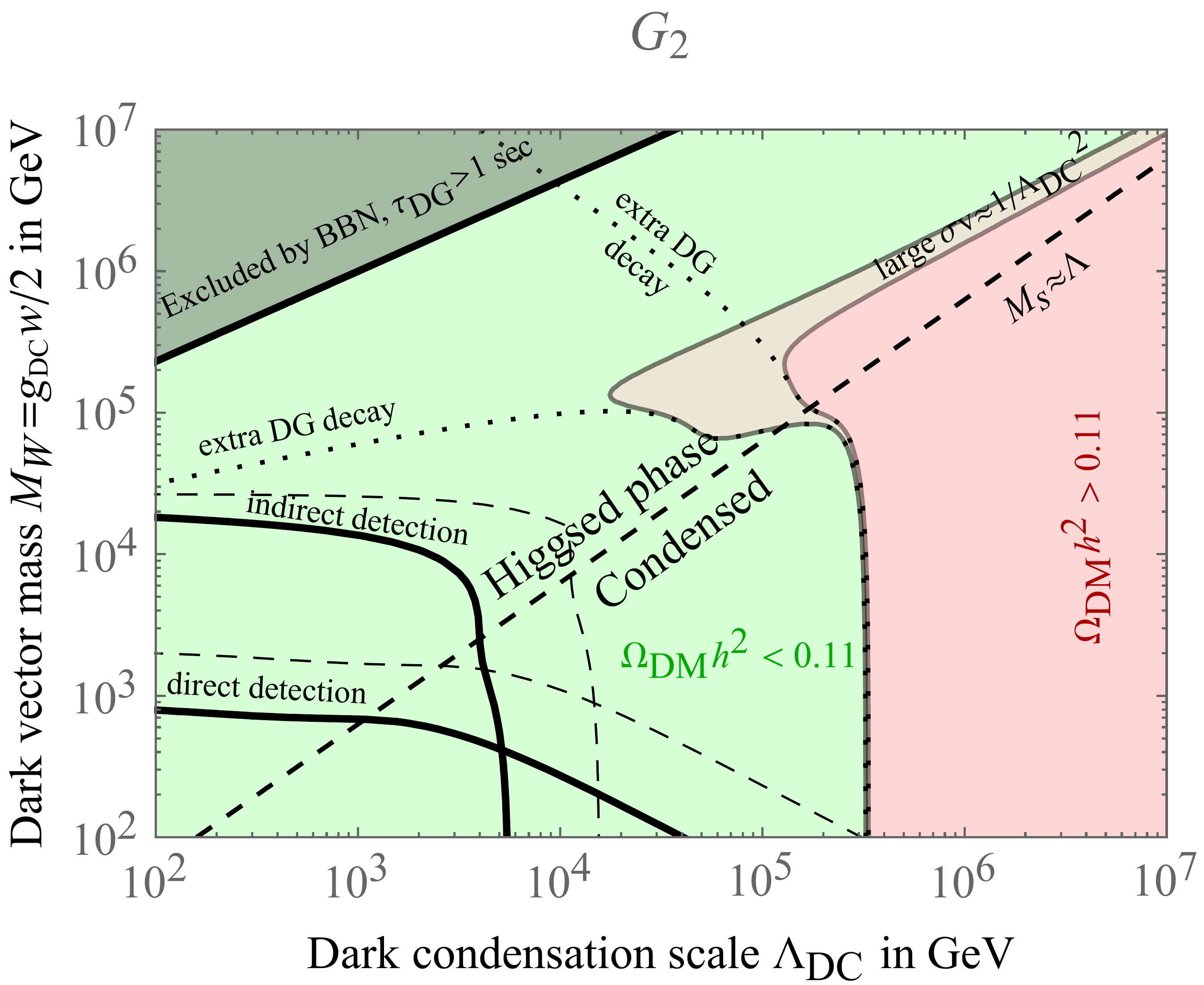}$$
\vspace{-1cm}
\caption{\em \label{fig:DMabundanceSpG2} 
As in fig.\fig{DMabundanceSU}, for $\Sp(\ndc)$ (left) and $G_2$ models (right).}
\end{figure}

The direct detection cross section is as in the SU(2) model~\cite{0811.0172}, and the region where DM is thermally produced is fully allowed, as shown in the left panel of fig.~\ref{fig:DMabundanceSpG2} and will be marginally probed in the future.

\medskip

\section{A fundamental of $G_2$}\label{sec:G2}
We consider $G_2$ because it
is the only exceptional group that is broken by its fundamental in an unique way, 
leaving the scalon and no extra scalars, such that the Higgs phase is expected to be equivalent to the condensed phase.
$G_2$ has 14 generators and a real
fundamental with dimension 7.
The invariant tensors of $G_2$ are $\delta_{ij}$, $\epsilon_{i_1\cdots i_7}$, and
$O_{ijk}$, the anti-symmetric tensor that defines
octonion multiplication $e_i e_j = -\delta_{ij}+O_{ijk}e_k$~\cite{Graves}.
The most generic renormalizable Lagrangian with a real scalar $\S_i$ in the fundamental
has one non-vanishing quartic $(\S\cdot\S)^2$,
while $\S^3$ cubic interactions vanish.
Such Lagrangian enjoys a $\S\to -\S$ accidental symmetry. 

\smallskip

Normalizing $G_2$ generators in the fundamental as $\Tr(T^a T^b)=\delta^{ab}$,
and generators in the adjoint as $\Tr(T^a T^b)=4\delta^{ab}$,
the  RGE are 
\beq 
(4\pi)^2 \frac{d\gDC}{d\ln\mu}  = -\frac{29}{2} \gDC^3,\qquad
(4\pi)^2 \frac{d\lambda_\S}{d\ln\mu} =4\gDC^4 -
24 \gDC^2 \lambda _\S +30 \lambda^{ 2} _\S .\eeq
\subsection{$G_2$: Higgs phase}
A vacuum expectation value of $\S$ breaks $G_2\to \SU(3)$.
The $G_2$ adjoint decomposes as ${ 14} = { 8}\oplus { 3} \oplus {\bar{3}} 
= \A \oplus \W \oplus \overline{\W}$ under $\SU(3)$. See Appendix \ref{app:genG2}
for the explicit embedding of the $\SU(3)$ subalgebra into $G_2$.
The perturbative spectrum is:
\begin{itemize}
\item the scalon $s$;
\item the 8 massless vectors $\A$ in the adjoint of $\SU(3)$;
\item vectors $\W$ with mass $M_\W^2 = \gDC^2w^2/3$  
in the ${ 3}\oplus {\bar 3}$ of SU(3). 
\end{itemize}
Notice that complex $\W$'s emerge from a real theory.
It is useful to compare $G_2$ to SU(4), that has 15 generators
and 8 scalar degrees of freedom in its complex fundamental {4}.
The $G_2\to\SU(3)$ theory differs from the $\SU(4)\to\SU(3)$ theory because of the absence
of the $\Z$ vector and because of the presence of $\epsilon^{\alpha\beta\gamma}\W_\alpha\W_\beta\W_\gamma$ gauge interactions.

\subsubsection*{Condensation of $\SU(3)$}
Taking into account the confinement of SU(3) gives the following singlets:
\begin{itemize}
\item mesons $\M=\W_\alpha\overline{ \W}^\alpha$ which decay to glue-balls;
\item $\A\A$ glue-balls;
\item baryons $\epsilon^{\alpha\beta\gamma}\W_\alpha\W_\beta\W_\gamma $ and $\epsilon_{\alpha\beta\gamma}\overline{\W}^\alpha \overline{\W}^\beta \overline{\W}^\gamma $  constructed contracting with the $\SU(3)$ anti-symmetric tensor $\epsilon$.
\end{itemize}
The $\W\W\W$ and the $ \overline{\W} \overline{\W} \overline{\W}$
decay to the same final state, in such a way that the decay amplitude cancels for an appropriate
combination $\Re\W\W\W=(\W\W\W + \overline{\W} \overline{\W} \overline{\W})/\sqrt{2} $, similarly to what happens for neutral kaons
(one combination is long lived, one combination is short lived).
Stability can be understood in terms of the charge conjugation symmetry C
of the Lagrangian:
\begin{equation}\label{eq:CG2}
\W_\alpha \to - \overline{\W}^\alpha, \qquad
 \A_{\rm imag} \to \A_{\rm imag} , \qquad \A_{\rm real} \to -\A_{\rm real}\,,
\end{equation}
where $\A_{\rm real} $ ($\A_{\rm imag}$) are the SU(3) vectors with real (imaginary) generators.
Interactions dictated by SU(3) gauge invariance respect this symmetry because it
reduces (up to a phase) to the usual SU(3) complex conjugation;
one can check that this symmetry is respected also by the 
extra $\W\W\W$, $\A\W\W\W$ interactions.
Then, the $\Re\W\W\W $ baryon is stable being odd under the C symmetry,
while  $\W\W\W - \overline{\W} \overline{\W} \overline{\W} $ is C-even and decays
through the $\W\W\W$ interactions.

\subsection{$G_2$: condensed phase}
In the confined $G_2$-invariant phase, the spectrum is described  by 
\begin{itemize}
\item $\M=\S_i\S_i$ mesons which decay in glue-balls;
\item $\G \G$ glue-balls;
\item $O_{ijk} \S_i\S_j\S_k$  baryons built with the rank-3 invariant antisymmetric tensor $O$
and with derivatives (not shown);
\item $\epsilon_{ijklmns} \S_i\G_{jk}\G_{lm}\G_{ns}$ 
baryons built with the $\epsilon$ invariant antisymmetric tensor.
The $G_2$ vectors $\G_{ij}=T^a_{ij}\G^a$  are anti-symmetric in $ij$  (like SO vectors)
and do not fill the most generic anti-symmetric matrix  (unlike SO vectors).

\end{itemize}
The two baryons have the same spin and mix~\cite{hep-lat/0302023}.
In the broken theory they are comparably heavy because
$\G\G\G$ gets its  $\A\W\W^*$ component.
The lightest baryon
is stable, because of the $\S\to-\S$ accidental symmetry. 

The symmetry that remains unbroken in the Higgs phase corresponds, in the condensed
phase, to $\S\to \eta \S$ where
$\eta = \diag(-1,-1,-1,1,1,1,1)$ in the basis given in Appendix~\ref{app:genG2}. This flips the ${ 3} + {\bar{3}}$ indices and leaves invariant the ${ 3} - {\bar{3}}$ indices as well as $\S_7$.
This component gets a vacuum expectation value in our basis, so that that
this symmetry remains unbroken in the Higgs phase. 
This symmetry acts on $G_2$ vectors as  $\G_{ij}\to(\eta \G\eta)_{ij}$ 
i.e.\
\begin{equation}
\G^{1,3,4,6,8,10,11,13} \to - \G^{1,3,4,6,8,10,11,13} ,\qquad \G^{2,5,7,9,12,14} \to \G^{2,5,7,9,12,14} .
\end{equation}
which agrees  with eq.~\eqref{eq:CG2}.
The compatibility of this symmetry with the Lie algebra is checked as follows: $f_{abc} = 0$ if one or three indices correspond to odd generators. 
This symmetry is an inner automorphism of the real group $G_2$ that
when restricted to its $\SU(3)$ subgroup acts as complex conjugation,
which is the outer automorphism of $\SU(3)$ that exchanges ${ 3} \leftrightarrow -{\bar{3}}$. 


Both $O\S\S\S$ and $\epsilon \S\G\G\G$ baryons are odd under $\eta$, because
both $\epsilon_{i_1\cdots i_7}$ and
$O_{ijk}$ contain an odd number of indices from the set $\{1,2,3\}$. Indeed, the only nonzero elements of the octonion algebra are
\be \label{eq:octonion}
O_{123} = O_{516} = O_{624} = O_{435} = O_{471} = O_{673} = O_{572} = 1
\ee
up to entries obtained by antisymmetry. Therefore, the two baryon structures have the same quantum numbers and mix into the physical stable baryon. To establish the correspondence of the baryons let us consider the Goldstone part of the baryon  $\Re(\epsilon_{\alpha \beta \gamma} \W_\alpha \W_\beta \W_\gamma)$ in the Higgs phase. For instance
\beq
\Re(\W_1 \W_2 \W_3) \sim \Re\(\frac{\S_1 + i \S_4}{\sqrt{2}} \frac{\S_2 + i \S_5}{\sqrt{2}} \frac{\S_3 + i \S_6}{\sqrt{2}}\)  
= \frac{\S_1 \S_2 \S_3 - \S_1 \S_5 \S_6 - \S_4 \S_2 \S_6 - \S_4 \S_5 \S_3}{2 \sqrt{2}} 
\eeq
coincides with the Goldstone part of $O_{ijk}\S_i\S_j\S_k$, by virtue of eq.~\eqref{eq:octonion}.

In conclusion, the same spectrum is obtained in the Higgs and condensed phases
of a $G_2$ gauge theory with a scalar in its fundamental.
The equivalence is more sophisticated because of the breaking of a real group to a complex subgroup:
\begin{itemize}
\item the meson $\S^T \S$ corresponds to the scalon $s$;
\item the operator $\S^T \D_\mu \S$ does not give rise to a $\Z_\mu$ due to the anti-symmetry of the generators (see footnote~\ref{foot:Z});

\item dark glue-balls in the condensed phase correspond to dark glue-balls in the Higgs phase;

\item the lightest baryon, admixture of $ \S \S \S$ and $ \S \G \G \G$, corresponds to the baryon $\Re \W \W \W $ of the Higgs phase; 

\item the C-even baryon $\Im \W\W\W$, that mixes with the scalon and with SU(3) glue-balls, corresponds to resonances of the mesons and glueballs of $G_2$.

\end{itemize}

\subsection{$G_2$: phenomenology}\label{PhenomenologyG2}
The theory is similar to the $\SU(4)\to \SU(3)$ theory, up to the absence of the $\Z$ boson
and to the presence of $\W\W\W$ interactions.
The perturbative $\W\W^*\to \A\A,\A s,ss$ DM annihilation cross sections are thereby 
equal to those given in eq.~(\ref{sys:sigmavrel}).
Furthermore there are extra $\W\W \to \W^*$ semi-annihilations,
as in DM models with an ad-hoc $\mathbb{Z}_3$ symmetry~\cite{1003.5912}.
The perturbative $\W$ relic density is thereby similar to the density in the $\SU(4)\to \SU(3)$ model.
$\W\W\W$ interactions give an extra difference at non-perturbative level:
when the $\W\W\W$ and $ \overline{\W} \overline{\W} \overline{\W}$ baryons form, 
only half of them survive in the stable C-odd component, analogously to a $K^0$ beam after the decay of the short-lived $K^0_S$.
As DM is now real, indirect detection is enhanced by a order one factor,
while direct detection is as in the $\SU(4) \to \SU(3)$ model
(after taking into account the slightly different RGE and thereby scalon mass).
Figure~\ref{fig:DMabundanceSpG2} (right) summarizes our final results.

\section{Conclusions}\label{concl}
We have studied models with a new dark gauge group $\G$ and a new dark scalar $\S$,
selected such that the Higgs phase (where $\S$ gets  a vacuum expectation value, breaking $\G$
to a sub-group $\H$) is dual to the confined phase (where $\G$ gets strongly interacting).
Fradkin, Shenker and others proved that this happens for $\G=\SU(2)$ with a scalar $\S$ in its fundamental.
We argued that the correspondence of the two phases holds whenever
the scalar $\S$ breaks $\G$ to a unique sub-group 
$\H$.\footnote{Indeed, in case the sub-group 
$\H$ were not unique, 
there would be different spectra of asymptotic 
states associated to each possible breaking.
Hence, the condensed phase of $\G$, 
which is presumably unique since it is dominated by 
gauge interactions, cannot be 
equivalent to the Higgs phases.}
In these cases $\S$ admits a single quartic self-coupling, and
the broken theory contains a single Higgs scalar, that we call $s$.
This happens when $\S$ fills a fundamental of the
$\SU(\ndc)$, $\SO(\ndc)$, $\Sp(\ndc)$, $G_2$ groups.
Table~\ref{tab:correspondence} summarizes how the Higgs/confinement duality is realized in each model. 
%
%

We studied such models from  the point of view of DM phenomenology.
When presenting final results, we further restricted the parameter space assuming that:
\begin{itemize}
\item the cosmological DM abundance is reproduced thermally;
\item the $\G\to \H$
symmetry breaking occurs dynamically {\em \`a la}  Coleman-Weinberg;
\item the $\S$ vacuum expectation value also induces the observed Higgs mass.
\end{itemize}
Thanks to these extra assumptions, DM phenomenology is described by one free parameter,
the dark gauge coupling $\gDC$ of $\G$. The confined phase is obtained smoothly for 
$\gDC\sim 4\pi / \sqrt{\ndc}$.
Smaller perturbative $\gDC$ correspond to the Higgs phase.
As strong interactions (either of $\G$ or $\H$) are often involved, 
cosmology often selects the DM mass typical of strong interactions: about $100\TeV$.
Such DM is heavy enough that the considered models are experimentally allowed.
Of course, some of the above assumptions can be relaxed, giving more general phenomenology. The various DM candidates are listed in table~\ref{tab:DM} for each case, together with their main features.

\begin{table}[t]
\begin{center}\footnotesize
\begin{tabular}{c|c|c}
Group & Higgs phase & Condensed phase \\
\hline
$\SU(\ndc) \to \SU(\ndc-1)$ & 
$
\begin{array} {lcl}
& s & \\ 
& \Z_\mu & \\
& \epsilon_{\ndc-1} \W^{\ndc-1} & \\
& \A \A & \\
& d\A\A\A & \\
\end{array}
$
& 
$
\begin{array} {lcl}
& \S^\dag \S & \\ 
& \S^\dag \D_\mu \S & \\
& \epsilon_{\ndc} \S^{\ndc} & \\
& \G \G& \\
& d\G\G\G & \\ 
\end{array}
$
\\
\hline
$\SO(\ndc) \to \SO(\ndc-1)$ & 
$
\begin{array} {lcl}
& s & \\ 
& \epsilon_{\ndc-1} \A \ldots \A \quad \text{(for odd $\ndc$)} & \\
& \epsilon_{\ndc-1} \W \A \ldots \A \quad \text{(for even $\ndc$)} & \\
& \A \A  & 
\end{array}
$
& 
$
\begin{array} {lcl}
& \S^T \S & \\ 
& \epsilon_{\ndc} \S \G \ldots \G \quad \text{(for odd $\ndc$)} & \\
& \epsilon_{\ndc} \G \ldots \G \quad \text{(for even $\ndc$)} & \\
& \G \G  & 
\end{array}
$
\\
\hline
$\Sp(\ndc) \to \Sp(\ndc-2)$ & 
$
\begin{array} {lcl}
& s, \X^\dag \X & \\ 
& \Z_\mu, \X^\dag \D_\mu \X & \\
& \W_\mu, \X^T \gamma_{\ndc-2} \D_\mu \X & \\
& \A \A  & 
\end{array}
$
& 
$
\begin{array} {lcl}
& \S^\dag \S & \\ 
& \S^\dag \D_\mu \S & \\
& \S^T \gamma_{\ndc} \D_\mu \S & \\
& \G \G  & 
\end{array}
$
\\
\hline
$G_2 \to \SU(3)$ & 
$
\begin{array} {lcl}
& s & \\ 
& \Re \W\W\W & \\
& \A \A  & 
\end{array}
$
& 
$
\begin{array} {lcl}
& \S^T \S & \\ 
& \S \S \S, \S \G \G \G & \\
& \G \G  & 
\end{array}
$ \\
\hline
\end{tabular}
\end{center}
\caption{\em\label{tab:correspondence}
Correspondence of the asymptotic states between the Higgs and confined phases.
} 
\end{table}

Our main results can be summarized as follows:
\begin{itemize}
\item In section \ref{sec:SU} we considered $\G=\SU(\NDC)$ with $\S$ in its complex fundamental representation.
In both phases the theory admits an unbroken accidental U(1) dark-baryon  number {and a charge conjugation in the dark sector}, leading to DM stability.
DM is composed by the baryon made by $\NDC$ scalars $\S$ and the C-odd dark glue-balls.
In the Higgs phase $\G$ is broken to $\H=\SU(\NDC-1)$, and one of the $\NDC$ scalars gets replaced by its vacuum expectation value $\langle\S\rangle$, so that DM is made by $\ndc-1$ heavy vectors {and by C-odd glue-balls}.
$\H$ confines at a lower scale $\LDC$,
giving a strong suppression of the cosmological DM relic density. 
Being made by heavy constituents, 
the size of heavy-vector DM (and thereby its cross sections in cosmology and in indirect detection)
is set by its Bohr-like radius.
Higgs soft theorems allowed to compute DM direct detection.
Fig.\fig{DMabundanceSU} summarizes DM phenomenology, showing  that
all experimental bounds are satisfied.
C-odd glue-balls are less important, as their contribution is comparable to
heavy vectors only if $\LDC \circa{>} M_\W$, a regime where we can only perform estimates.
Furthermore, DM is accompanied by lighter, unstable C-even dark glue-balls that can {potentially} be probed by their coupling to the Higgs.

\item In section~\ref{sec:SO} we considered $\G=\SO(\NDC)$ with $\S$ in its real fundamental representation.
DM is  stable because of O-parity, a symmetry related to SO groups analogous to how
parity is related to the rotation group.
For odd $\NDC$, DM is the baryon made by one heavy scalar $\S$.
For even $\NDC$, DM is an odd dark glue-ball containing no scalars $\S$.
In the Higgs phase $\G$ is broken to $\H=\SO(\NDC-1)$, that confines at a lower scale $\LDC$,
giving  a strong suppression of the cosmological DM relic density.
Since DM contains light dark gluons, it has a larger size set by $1/\LDC$. 
An extension of  Higgs soft theorems allowed to compute direct detection of odd-ball DM.
Fig.\fig{DMabundanceSO} summarizes DM phenomenology, showing  that
all experimental bounds are satisfied.

\begin{table}
\begin{tabular}{|c|c|l|c|c}
Group & Global symmetry & {\qquad\quad DM candidate} & DM Annihilation\\
\hline
$\SU(\ndc)$ & Dark baryon number & Baryon $\epsilon S^\ndc \cong \W^{n}$ & Bohr-like -- $1/\LDC^2$\\
& Charge conjugation & Glue-balls $d\G\G\G\cong d\A\A\A$ & $1/\LDC^2$\\
$\SO(\ndc_{\rm even})$& O-parity & 1-ball $\epsilon \G^{\ndc/2} \cong \W \A^{(n-1)/2}$ &  $1/\LDC^{2}$\\
$\SO(\ndc_{\rm odd})$& O-parity & 0-ball $\epsilon \S\G^{(\ndc-1)/2} \cong \A^{n/2}$ & $1/\LDC^{2}$\\
$\Sp(\ndc)$ & Dark baryon number & Meson $\S\gamma\S \cong \W,\X\X$ & Perturbative\\
$G_2$ & Inner automorphism & Baryon $\epsilon\S\G^3,O\S^3 \cong {\rm Re}\,\W^3$ & Bohr-like -- $1/\LDC^2$\\
\end{tabular}
\caption{\em Dark matter candidates, thier stabilizing symmetries, and dominant annihilation mechanisms in the various models.\label{tab:DM}
}
\end{table}

\item In section~\ref{sec:Sp} we considered $\G=\Sp(\NDC)$ with $\S$ in its pseudo-real fundamental.
DM is stable thanks to an accidental U(1) dark baryon number.
In the Higgs phase $\G$ is broken to $\H=\Sp(\NDC-2)$, giving two co-stable vector
DM candidates $\W$ (neutral under $\H$ and with dark baryon number 2) and $\X$ (charged under $\H$ and with dark baryon number 1), with masses $M_\X= M_\W/\sqrt{2}$.
When $\H$ confines at a lower scale $\LDC$, two $\X$'s form a meson and
their cosmological DM relic density gets strongly suppressed.
DM remains as $\W$ with  cosmological relic density approximately not suppressed by $\H$ confinement.
Because of this, dark glue-balls can be especially light in Sp models.
Up to the presence of dark glue-balls, DM phenomenology is similar to the $\SU(2)=\Sp(2)$ model.
Fig.\fig{DMabundanceSpG2} (left) summarizes DM phenomenology, showing  that
all experimental bounds are satisfied.

\item In section~\ref{sec:G2} we considered the exceptional group $G_2$ with $\S$ in its real fundamental.
In the confined phase, the $\S\S\S$ and $\S\G\G\G$ baryons remain stable
thanks to an accidental $\S\to-\S$ symmetry.
In the Higgs phase $G_2$ is broken to $\H=\SU(3)$ and the theory contains massive
$\W \oplus\overline{\W}$ vectors in the ${ 3}\oplus{\bar 3}$.
The theory contains $\W\W\W$ gauge interaction characteristic of $G_2$, which give $\W\W\to \W^*$
processes.
As a result the $\Im\W\W\W$ baryon decays, while the
$\Re\W\W\W$ remains as a stable DM candidate, thanks to 
an inner automorphism of $G_2$ that reduces to charge conjugation of $\SU(3)$.
Stability arises as a quantum mechanical interference phenomenon, 
analogous to how the neutral kaons split into long-lived and short-lived eigenstates.
DM size is set by the Bohr-like radius.
Fig.~\ref{fig:DMabundanceSpG2} (right) summarizes DM phenomenology, showing  that
all experimental bounds are satisfied.
\end{itemize}
As we sometimes relied on approximations,
various aspects of each model can be more precisely computed.
Furthermore, it will be interesting to see if other choices of scalar representations
that do not satisfy the Higgs/confinement duality  lead to  DM candidates with distinct phenomenology.


\footnotesize

\subsubsection*{Acknowledgements}
This work was supported by the ERC grant NEO-NAT, by MIUR under contract number 2017L5W2PT, and by the INFN grant FLAVOR.
We thank Claudio Bonati, Christian Gross, Thomas Hambye,
Paolo Panci, Michele Redi, and Filippo Sala for discussions.
 
\appendix

\section{Generators}
\label{app:gen}

For completeness we provide here the 
$\SU(\ndc)$, $\SO(\ndc)$, $\Sp(\ndc)$ and $G_2$ 
generators in the fundamental representation.

\subsection{$\SU(\ndc)$}
\label{app:genSU}

The $\SU(\ndc)$ generators are given in terms of a 
generalization of the Pauli matrices $\sigma_a/2$:
\begin{eqnsystem}{sys:TSU}
\label{eq:T1ab}
(T^{(1)}_{\alpha\beta})_{\gamma\delta} &= \frac{1}{2} (\delta_{\alpha\gamma} \delta_{\beta\delta} + \delta_{\alpha\delta} \delta_{\beta\gamma})  \quad (1 \leq \alpha < \beta \leq \ndc) \\
\label{eq:T2ab}
(T^{(2)}_{\alpha\beta})_{\gamma\delta} &= -\frac{i}{2} (\delta_{\alpha\gamma} \delta_{\beta\delta} - \delta_{\alpha\delta} \delta_{\beta\gamma}) \quad (1 \leq \alpha < \beta \leq \ndc) \\
\label{eq:T3ab}
(T^{(3)}_{\alpha})_{\gamma\delta} 
&= 
\left\{
\begin{array}{cl}
\frac{1}{\sqrt{2\alpha(\alpha-1)}} \delta_{\gamma\delta}  & (\gamma < \alpha) \\
-\sqrt{\frac{\alpha-1}{2\alpha}} \delta_{\gamma\delta}  &  (\gamma = \alpha \quad 2 \leq \alpha  \leq \ndc) \\
0 &  (\gamma > \alpha) 
\end{array}
\right. \, .
\end{eqnsystem}
Altogether they are $\frac{1}{2}\ndc(\ndc-1) + \frac{1}{2}\ndc(\ndc-1) + (\ndc-1) = \ndc^2 -1$ 
generators, which can be collected as 
\begin{align}
T^1 &= T^{(1)}_{12} \, , \qquad 
T^2 = T^{(2)}_{12} \, , \qquad 
T^3 = T^{(3)}_2 \, , \qquad 
T^4 = T^{(1)}_{13} \, , \qquad 
T^5 = T^{(2)}_{13} \, , \qquad \\
T^6 &= T^{(1)}_{23} \, , \qquad 
T^7 = T^{(2)}_{23} \, , \qquad 
T^8 = T^{(3)}_{3} \, , \qquad \ \ \ \;
\ldots \qquad \ \,
T^{\ndc^2-1} = T^{(3)}_{\ndc} \, ,  
\end{align}
with normalization $\Tr(T^a T^b) = \frac{1}{2} \delta^{ab}$. 

\subsection{$\SO(\ndc)$}
\label{app:genSO}

The $\frac{1}{2}\ndc(\ndc-1)$ generators of $\SO(\ndc)$ are given in terms of the
$T^{(2)}_{\alpha\beta}$ SU generators
defined in eq.~(\ref{eq:T2ab}) as 
\beq
T^1 = 2T^{(2)}_{12} \, , \qquad 
T^2 = 2T^{(2)}_{13} \, , \qquad 
\ldots \qquad  
T^{\ndc -1} = 2T^{(2)}_{1\ndc} \, , \qquad 
\ldots \qquad  
T^{\frac{1}{2}\ndc(\ndc-1)} = 2T^{(2)}_{\ndc-1,\ndc} \, ,
\eeq
with normalization $\Tr(T^a T^b) = 2 \delta^{ab}$.

\subsection{$\Sp(\ndc)$}
\label{app:genSp}
Symplectic Lie groups exist for even $\ndc = 2 \ell$. 
The $\ell(2\ell+1)$ generators of $\Sp(2\ell)$ can be written 
in terms of the $\ell$-dimensional SU generators $T^{(1)}_{\alpha\beta}$ and 
$T^{(2)}_{\alpha\beta}$ defined in eq.~(\ref{sys:TSU}) as
\beq \label{eq:genSpN}
\frac{1}{\sqrt{2}} T^{(2)}_{\alpha\beta} \otimes \One_2 \, , \qquad 
\frac{1}{\sqrt{2}} T^{(1)}_{\alpha\beta} \otimes \sigma_k \, , \qquad 
\frac{1}{2} T^{(1)}_{\alpha\alpha} \otimes \sigma_k \, ,
\eeq
for $1\leq \alpha < \beta \leq \ell$ and $k=1,2,3$. 
In fact, these are $\frac{1}{2}\ell(\ell-1) + \frac{1}{2}\ell(\ell-1) \cdot 3 + \ell \cdot 3 = \ell(2\ell+1)$ elements,  
which can be collected as 
\begin{eqnsystem}{sys:Spgen}
T^1 &= \frac{1}{\sqrt{2}} T^{(2)}_{12} \otimes \mathbb{1}_2 
\qquad \dots \qquad 
T^{\frac{1}{2}\ell(\ell-1)} = \frac{1}{\sqrt{2}} T^{(2)}_{\ell-1,\ell} \otimes \mathbb{1}_2 
 \\
T^{\frac{1}{2}\ell(\ell-1)+1} &= \frac{1}{\sqrt{2}} T^{(1)}_{12} \otimes \sigma_1 
\qquad \dots \qquad
T^{2\ell(\ell-1)} = \frac{1}{\sqrt{2}} T^{(1)}_{\ell-1,\ell} \otimes \sigma_3 \\
T^{2\ell(\ell-1)+1} &= \frac{1}{2} T^{(1)}_{11} \otimes \sigma_1 
\qquad \dots \qquad
T^{\ell(2\ell+1)} = \frac{1}{2} T^{(1)}_{\ell\ell} \otimes \sigma_3 \, , 
\end{eqnsystem}
with normalization $\Tr(T^a T^b) = \frac{1}{2} \delta^{ab}$. 
The invariant tensor 
$\gamma_\ndc \equiv \One_{\ndc/2}\otimes i\sigma_2$
satisfies  $(T^a)^T \gamma_\ndc + \gamma_\ndc T^a = 0$.

\subsection{$G_2$}
\label{app:genG2}
 
$G_2$ has 14 generators which can be written in terms of the 7-dimensional matrices $T^{(2)}_{\alpha\beta}$ defined in eq.~(\ref{eq:T2ab}) as \cite{Gunaydin:1973rs}
\begin{eqnsystem} {sys:G2gen}
T^1 &= T^{(2)}_{51} - T^{(2)}_{24} \qquad T^{8} = \tfrac{1}{\sqrt{3}} (T^{(2)}_{24} + T^{(2)}_{51} - 2 T^{(2)}_{73}) \\
T^2 &= T^{(2)}_{54} - T^{(2)}_{12} \qquad T^{9} = -\tfrac{1}{\sqrt{3}} (T^{(2)}_{54} + T^{(2)}_{12} - 2 T^{(2)}_{67}) \\
T^3 &= T^{(2)}_{25} - T^{(2)}_{14} \qquad T^{10} = \tfrac{1}{\sqrt{3}} (T^{(2)}_{14} + T^{(2)}_{25} - 2 T^{(2)}_{36}) \\
T^4 &= T^{(2)}_{43} - T^{(2)}_{16} \qquad T^{11} = \tfrac{1}{\sqrt{3}} (T^{(2)}_{16} + T^{(2)}_{43} - 2 T^{(2)}_{72}) \\
T^5 &= T^{(2)}_{31} - T^{(2)}_{46} \qquad T^{12} = \tfrac{1}{\sqrt{3}} (T^{(2)}_{46} + T^{(2)}_{31} - 2 T^{(2)}_{57})  \\
T^6 &= T^{(2)}_{62} - T^{(2)}_{35} \qquad T^{13} = \tfrac{1}{\sqrt{3}} (T^{(2)}_{35} + T^{(2)}_{62} - 2 T^{(2)}_{71})  \\
T^7 &= T^{(2)}_{65} - T^{(2)}_{23} \qquad T^{14} = -\tfrac{1}{\sqrt{3}} (T^{(2)}_{65} + T^{(2)}_{23} - 2 T^{(2)}_{47}) \, ,
\end{eqnsystem} 
with normalization $\Tr(T^a T^b) = \delta^{ab}$. The adjoint decomposes under $\SU(3)$ as ${ 14} = { 8} \oplus { 3} \oplus {\bar{3}} = \A \oplus \W \oplus \overline{\W}$. Among the 7 dimensions, the first three correspond to the embedding of ${ 3} \oplus {\bar 3}$, the second three to ${ 3} - {\bar 3}$, and the 7th one to the singlet direction.  The $\SU(3)$ subalgebra is spanned by $\{ T^1,\ldots,T^7,-T^{10} \}$, with the $\SU(3)$ adjoint vectors embedded as $\A_{1,\ldots,7} = \G_{1,\ldots,7}$, $\A_8 = - \G_{10}$. The $\W$ are embedded as  
\begin{equation}
\W + \overline \W = (\G_8, \G_{11}, \G_{13}) , \quad  i(\W - \overline \W) = (\G_9, -\G_{12}, \G_{14}) \;.
\end{equation}

\section{Feynman rules} 
\label{app:SUNFeyn}

To derive the Feynman rules for the $\SU(\ndc) \to \SU(\ndc-1)$ breaking pattern 
we decompose the Lagrangian under the unbroken $\SU(\ndc-1)$ as follows
\begin{align}
-\frac{{\G}^a_{\mu\nu}{\G}^{a\,\mu\nu}}{4}+|\D_\mu\S|^2 
 &= \nonumber 
-\frac{1}{4} \A^a_{\mu\nu} \A^{a\mu\nu} -\frac{1}{2} \W^\dag_{\mu\nu} \W^{\mu\nu} 
-\frac{1}{4} \Z_{\mu\nu} \Z^{\mu\nu} 
- i \gDC 
(\W^\dag_{\mu} T^a_{\ndc -1} \W_{\nu}) \A^{a\mu\nu} \nonumber \\
&
+ i\gDC f_\ndc^{\ndc^2-1} \((\partial_{[\mu} \W_{\nu]}^\dag)\W^\nu \Z^\mu - \W_\nu^\dag (\partial^{[\mu}\W^{\nu]})\Z_\mu - \W_{[\mu}^\dag \W_{\nu]} \partial^\mu \Z^\nu)\) 
\nonumber \\
&-\frac{\gDC^2}{2} \ndc \( \W^\dag_\mu \W^\mu \Z_\nu \Z^\nu - \W^\dag_\mu \W_\nu \Z^\mu \Z^\nu \)
\nonumber\\
&- \gDC^2 f_\ndc^{\ndc^2-1} \( 2 \Z^\mu \A_\mu^a (\W_\nu^\dag T_{\ndc-1}^a \W^\nu) - (\Z_\mu \A_\nu^a + \Z_\nu \A_\mu^a) (\W^{\dag\mu} T_{\ndc-1}^a \W^\nu)\)
\nonumber \\
&+ \frac{1}{2} \partial_\mu s \partial^\mu s 
+ M^2_\W (1+ \frac{s}{w})^2 \W^\dag_\mu \W^\mu
+ \frac{1}{2} M^2_\Z (1+ \frac{s}{w})^2 \Z_\mu \Z^\mu +\cdots
\end{align}
where $\cdots$ denotes $\W\W^*\W\W^*$ vertices. We defined 
$f_\ndc^{\ndc^2-1} = \sqrt{\ndc/(2(\ndc-1))}$, 
$\D_\mu = \partial_\mu - i \gDC T^a_{\ndc -1} \A^a_\mu$
and
\beq 
\W_{\mu\nu} = \D_\mu \W_\nu - \D_\nu \W_\mu,\qquad
\Z_{\mu\nu} = \partial_\mu \Z_\nu - \partial_\nu \Z_\mu,\qquad
\A^a_{\mu\nu} = \partial_\mu \A^a_\nu - \partial_\nu \A^a_\mu + \gDC f^{abc}_{\ndc -1} \A^b_\mu \A^b_\nu \,.
\eeq
The Feynman vertices with all momenta $p_i$ incoming are: 
\begin{itemize}
\item[] \hspace{-1cm} 
\begin{minipage}{0.2\textwidth}
\includegraphics[width=1.\textwidth]{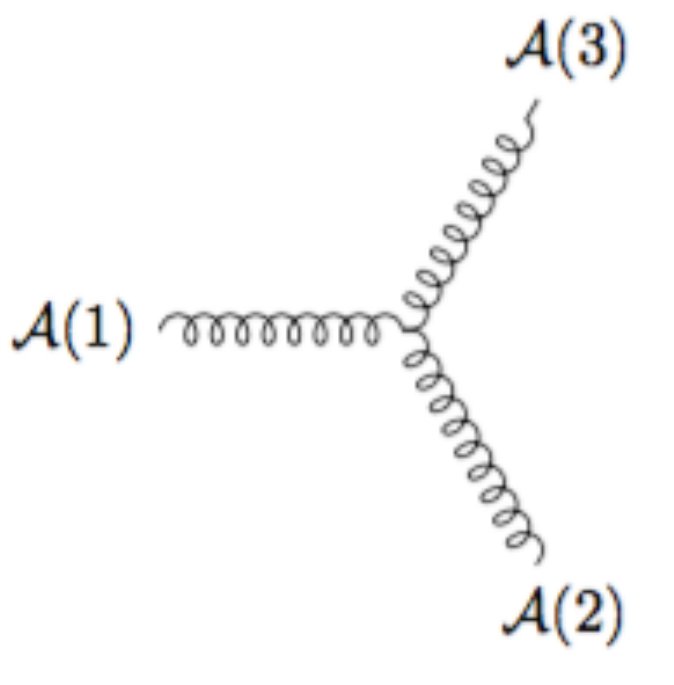}
\end{minipage}
\begin{minipage}{0.5\textwidth}
\vspace{-0.8cm}
\beq 
= \ \gDC f^{a_1a_2a_3}_{\ndc-1} \[ 
g_{\mu_1 \mu_2} (p_1^{\mu_3}-p_2^{\mu_3}) 
- g_{\mu_1 \mu_3} (p_1^{\mu_2}-p_3^{\mu_2}) 
+ g_{\mu_2 \mu_3} (p_2^{\mu_1}-p_3^{\mu_1})
\] \nonumber
\eeq
\end{minipage}
\item[] \hspace{-1cm} 
\begin{minipage}{0.2\textwidth}
\includegraphics[width=1.\textwidth]{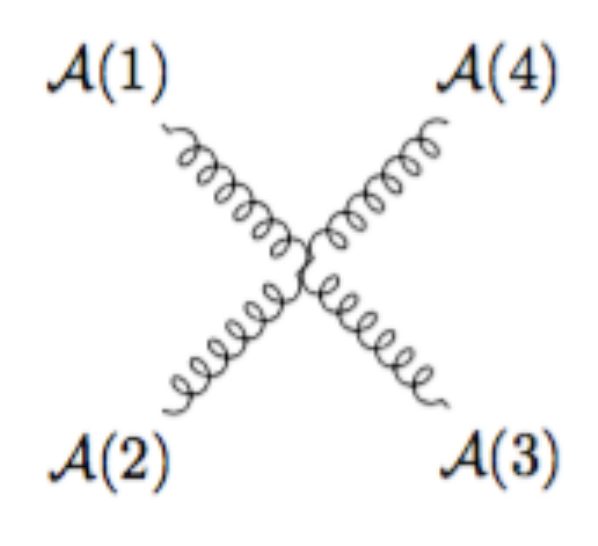}
\end{minipage}
\begin{minipage}{0.5\textwidth}
\begin{align} 
= \ i \gDC^2 
&\[
f^{a_1a_3c}_{\ndc-1} f^{a_2a_4c}_{\ndc-1} 
(g_{\mu_1 \mu_4} g_{\mu_2 \mu_3} - g_{\mu_1 \mu_2} g_{\mu_3 \mu_4}) \right. \nonumber \\
& + f^{a_1a_2c}_{\ndc-1} f^{a_3a_4c}_{\ndc-1} 
(g_{\mu_1 \mu_4} g_{\mu_2 \mu_3} - g_{\mu_1 \mu_3} g_{\mu_2 \mu_4}) \nonumber \\
& \left. + f^{a_1a_4c}_{\ndc-1} f^{a_2a_3c}_{\ndc-1} 
(g_{\mu_1 \mu_3} g_{\mu_2 \mu_4} - g_{\mu_1 \mu_2} g_{\mu_3 \mu_4}) \] \nonumber
\end{align}
\end{minipage}
\item[] \hspace{-1cm} 
\begin{minipage}{0.2\textwidth}
\includegraphics[width=1.\textwidth]{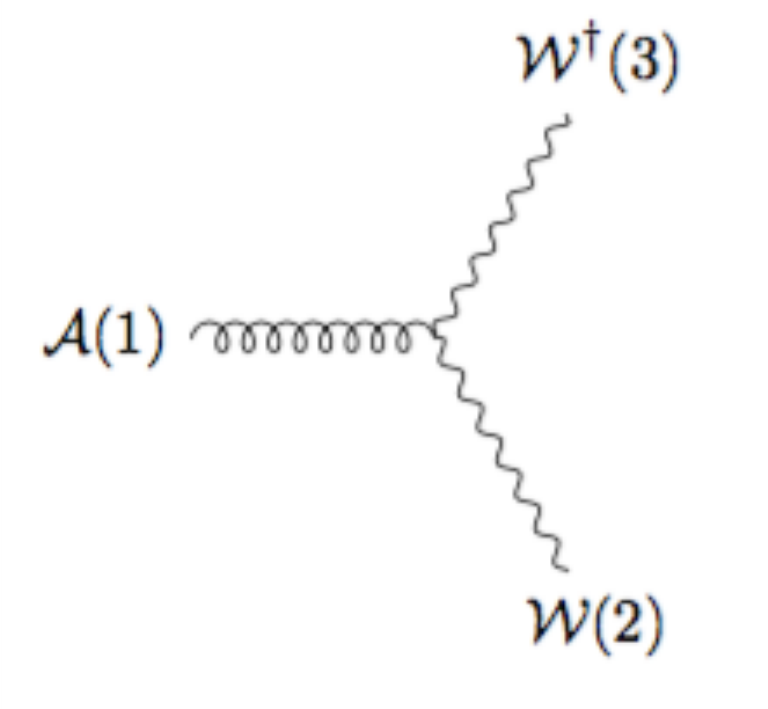}
\end{minipage}
\begin{minipage}{0.5\textwidth}
\begin{align}
= \ i \gDC (T^{a_1}_{\ndc-1})_{m_3 m_2} 
&\[ 
g_{\mu_1\mu_2} p_2^{\mu_3} 
- g_{\mu_1\mu_3} p_3^{\mu_2} 
- g_{\mu_2\mu_3} p_2^{\mu_1} 
+ g_{\mu_2\mu_3} p_3^{\mu_1} 
\right. \nonumber \\ 
&\left. - p_1^{\mu_3} g_{\mu_1\mu_2} 
+ p_1^{\mu_2} g_{\mu_1\mu_3} 
\] \nonumber
\end{align}
\end{minipage}
\item[] \hspace{-1cm} 
\begin{minipage}{0.2\textwidth}
\includegraphics[width=1.\textwidth]{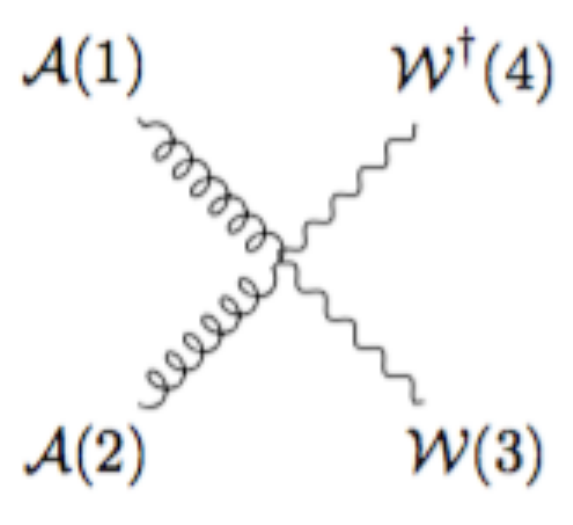}
\end{minipage}
\begin{minipage}{0.5\textwidth}
\begin{align} 
= \ i \gDC^2 
&\[ 
(T^{a_1}_{\ndc-1} T^{a_2}_{\ndc-1})_{m_4 m_3} 
\(g_{\mu_1\mu_3} g_{\mu_2\mu_4} 
- g_{\mu_1\mu_2} g_{\mu_3\mu_4}\) \right. \nonumber \\
& + (T^{a_2}_{\ndc-1} T^{a_1}_{\ndc-1})_{m_4 m_3}  \( 
g_{\mu_1\mu_4} g_{\mu_2\mu_3}
- g_{\mu_1\mu_2} g_{\mu_3\mu_4} \) \nonumber \\
& \left. -i f^{a_1a_2c}_{\ndc-1} (T^{c}_{\ndc-1})_{m_4 m_3} \( g_{\mu_1\mu_4} g_{\mu_2\mu_3} 
- g_{\mu_1\mu_3} g_{\mu_2\mu_4} \)
\] \nonumber 
\end{align}
\end{minipage}
\item[] \hspace{-1cm} 
\begin{minipage}{0.2\textwidth}
\includegraphics[width=1.\textwidth]{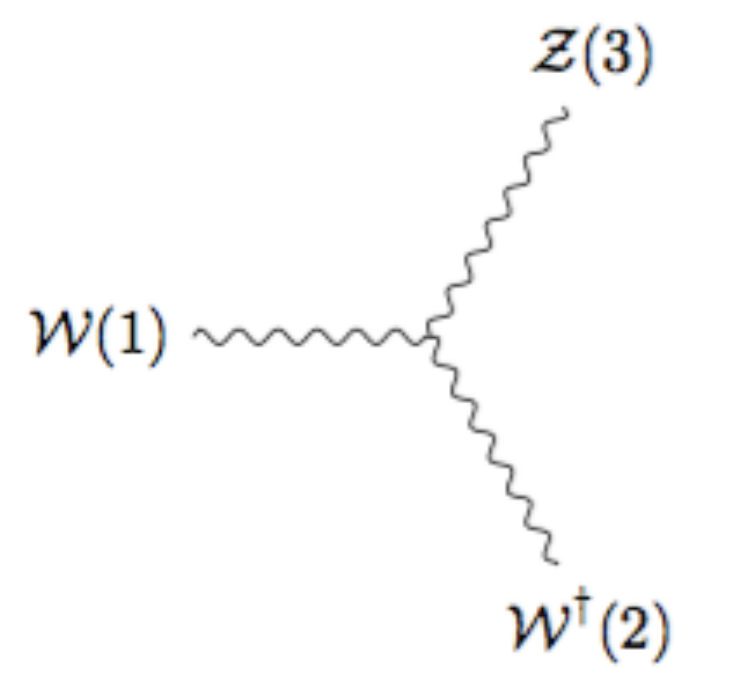}
\end{minipage}
\begin{minipage}{0.5\textwidth}
\vspace{-0.8cm}
\beq 
= \ - i \gDC f_\ndc^{\ndc^2-1} \[ 
g_{\mu_1 \mu_2} (p_1^{\mu_3}-p_2^{\mu_3}) 
- g_{\mu_1 \mu_3} (p_1^{\mu_2}-p_3^{\mu_2}) 
+ g_{\mu_2 \mu_3} (p_2^{\mu_1}-p_3^{\mu_1})
\] \nonumber 
\eeq
\end{minipage}
\item[] \hspace{-1cm} 
\begin{minipage}{0.2\textwidth}
\includegraphics[width=1.\textwidth]{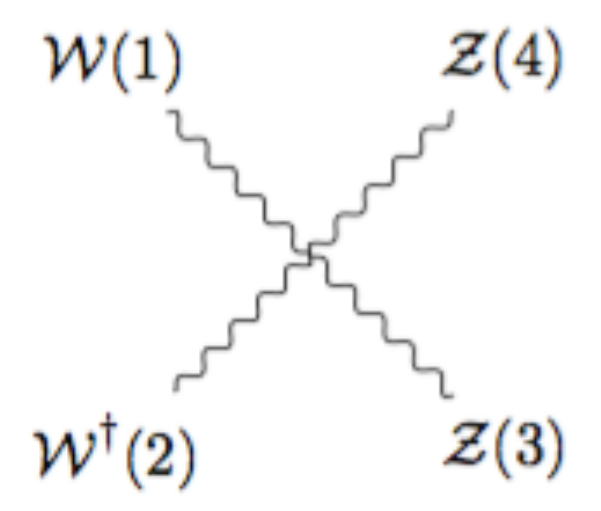}
\end{minipage}
\begin{minipage}{0.5\textwidth}
\vspace{-0.8cm}
\beq 
= \ -i \frac{\gDC^2}{2} \ndc \( 
2 g_{\mu_1\mu_2} g_{\mu_3\mu_4} - g_{\mu_1\mu_4} g_{\mu_2\mu_3}
- g_{\mu_1\mu_3} g_{\mu_2\mu_4}
\)(\delta_{\ndc-1})_{m_1 m_2} \nonumber 
\eeq
\end{minipage}
\item[] \hspace{-1cm} 
\begin{minipage}{0.2\textwidth}
\includegraphics[width=1.\textwidth]{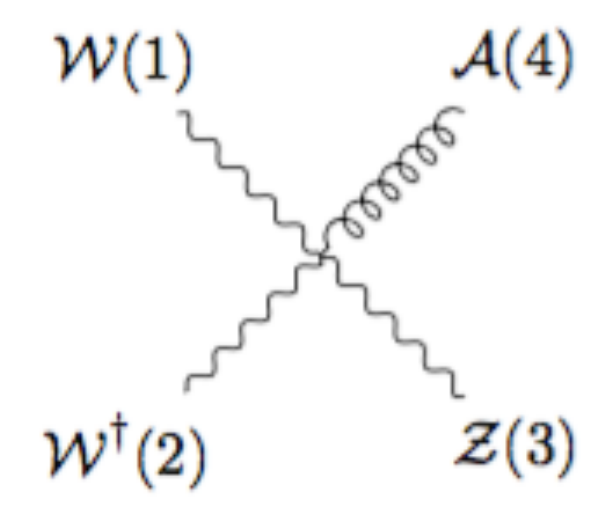}
\end{minipage}
\begin{minipage}{0.5\textwidth}
\vspace{-0.8cm}
\beq 
= \ -i \gDC^2 f^{\ndc^2-1}_{\ndc} \( 
2 g_{\mu_1\mu_2} g_{\mu_3\mu_4} - g_{\mu_1\mu_4} g_{\mu_2\mu_3}
- g_{\mu_1\mu_3} g_{\mu_2\mu_4}
\)(T_{\ndc-1}^{a_4})_{m_2 m_1} \nonumber 
\eeq
\end{minipage}
\item[] \hspace{-1cm} 
\begin{minipage}{0.2\textwidth}
\includegraphics[width=1.\textwidth]{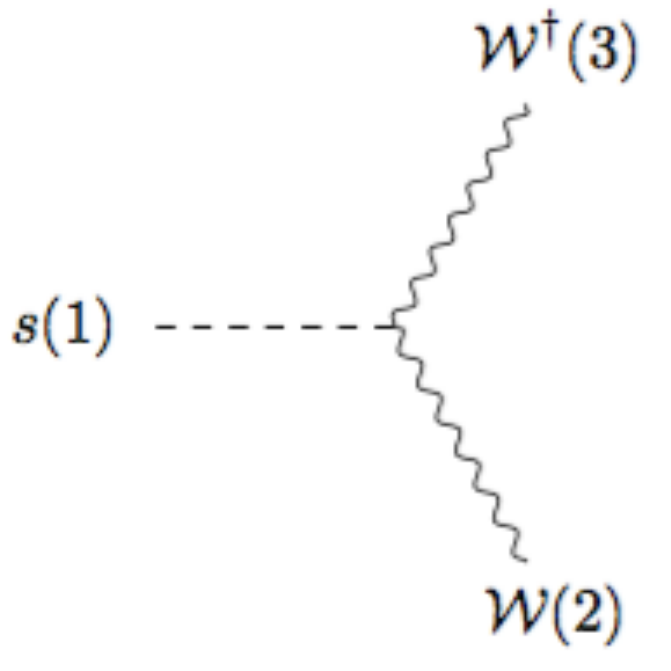}
\end{minipage}
\begin{minipage}{0.2\textwidth}
\vspace{-0.8cm}
\beq 
= \ 2 i \frac{M^2_\W}{w} (\delta_{\ndc-1})_{m_2m_3} g_{\mu_2\mu_3} \nonumber 
\eeq
\end{minipage}
\item[] \hspace{-1cm} 
\begin{minipage}{0.2\textwidth}
\includegraphics[width=1.\textwidth]{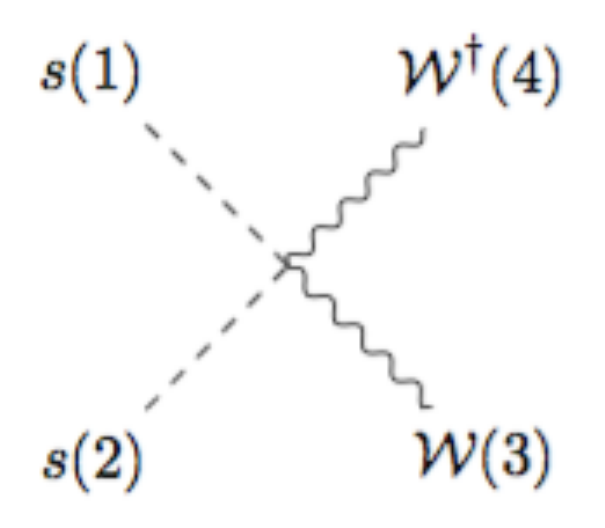}
\end{minipage}
\begin{minipage}{0.2\textwidth}
\vspace{-0.65cm}
\beq 
= \ 2 i \frac{M^2_\W}{w^2} (\delta_{\ndc-1})_{m_3m_4} g_{\mu_3\mu_4} \nonumber 
\eeq
\end{minipage}
\item[] \hspace{-1cm} 
\begin{minipage}{0.2\textwidth}
\includegraphics[width=1.\textwidth]{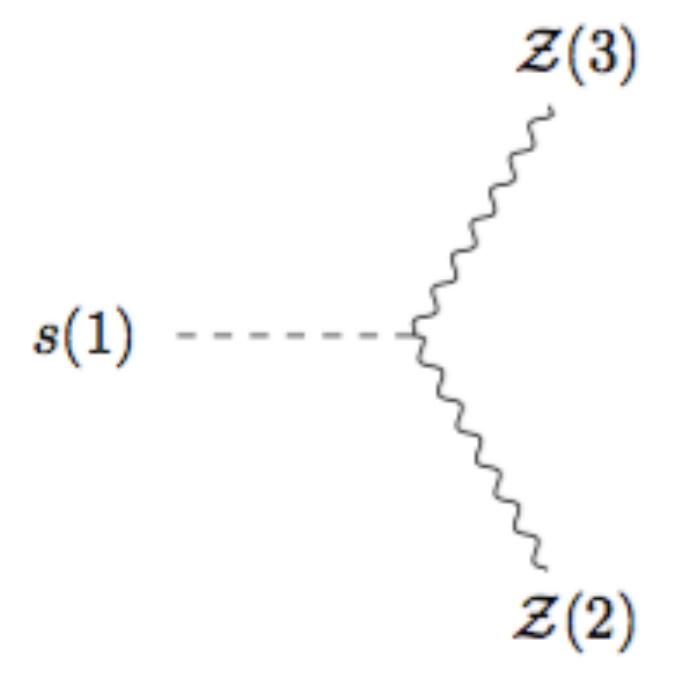}
\end{minipage}
\begin{minipage}{0.2\textwidth}
\vspace{-0.8cm}
\beq 
= \ 2 i \frac{M^2_\Z}{w} g_{\mu_2\mu_3} \nonumber 
\eeq
\end{minipage}
\item[] \hspace{-1cm} 
\begin{minipage}{0.2\textwidth}
\includegraphics[width=1.\textwidth]{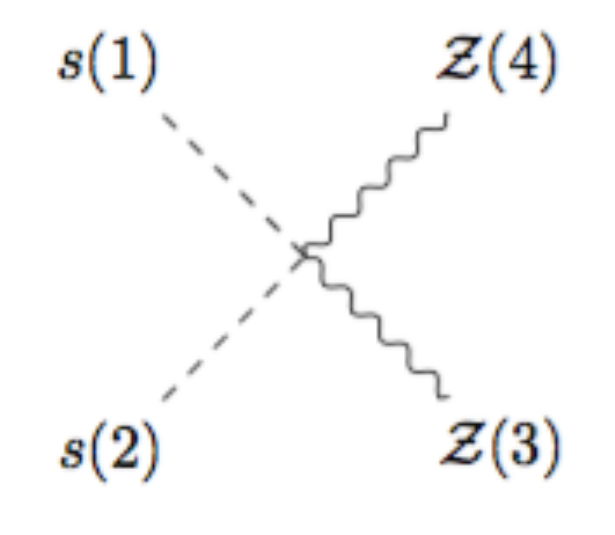}
\end{minipage}
\begin{minipage}{0.2\textwidth}
\vspace{-0.8cm}
\beq 
= \ 2 i \frac{M^2_\Z}{w^2} g_{\mu_3\mu_4} \nonumber 
\eeq
\end{minipage}
\end{itemize}
Similar expressions can be derived for the other groups
considered in this paper. E.g.~the $\SO(\ndc) \to \SO(\ndc-1)$ case
is simply obtained  from $\SU(\ndc) \to \SU(\ndc-1)$ by 
dropping vertices involving $\Z$ bosons
and taking into account that $\W$'s become real.
The $G_2$ group gives extra $\W\W\W$ interactions, and Sp gives extra vectors $\X$.


\begin{thebibliography}{nnn}\bibitem{ScalarSinglet}
\article[Silveira:1985rk]{V. Silveira, A. Zee}{Phys. Lett.}{161B}{136}{1985}{Scalar phantoms}.
\article[hep-ph/0011335]{C.P. Burgess, M. Pospelov, T. ter Veldhuis}{Nucl. Phys.}{B619}{709}{2000}
{The Minimal model of nonbaryonic dark matter: A Singlet scalar}.
\article[0912.5038]{M. Farina, D. Pappadopulo, A. Strumia}{Phys. Lett.}{B688}{329}{2009}
{CDMS stands for Constrained Dark Matter Singlet}.
\article[1306.4710]{J.M. Cline, K. Kainulainen, P. Scott, C. Weniger}{Phys. Rev.}{D88}{055025}{2013}
{Update on scalar singlet dark matter}.
\heparticle[1705.07931]{GAMBIT Collaboration}{Status of the scalar singlet dark matter model}.


\bibitem{1111.4482}
\article[1111.4482]{O. Lebedev, H.M. Lee, Y. Mambrini}{Phys. Lett.}{B707}{570}{2012}
{Vector Higgs-portal dark matter and the invisible Higgs}.


\bibitem{0811.0172}
\article[0811.0172]{T. Hambye}{JHEP}{0901}{028}{2008}
{Hidden vector dark matter}.


\bibitem{0907.1007}
\article[0907.1007]{T. Hambye, M.H.G. Tytgat}{Phys. Lett.}{B683}{39}{2009}{Confined hidden vector dark matter}.


\bibitem{1306.2329}
\article[1306.2329]{T. Hambye, A. Strumia}{Phys. Rev.}{D88}{055022}{2013}{Dynamical generation of the weak and Dark Matter scale}.


\bibitem{Osterwalder:1977pc}
\article[Osterwalder:1977pc]{K. Osterwalder, E. Seiler}{Annals Phys.}{110}{440}{1978}
{Gauge Field Theories on the Lattice}.


\bibitem{Fradkin:1978dv}
\article[Fradkin:1978dv]{E.H. Fradkin, S.H. Shenker}{Phys. Rev.}{D19}{3682}{1979}
{Phase Diagrams of Lattice Gauge Theories with Higgs Fields}.


\bibitem{Banks:1979fi}
\article[Banks:1979fi]{T. Banks, E. Rabinovici}{Nucl. Phys.}{B160}{349}{1979}
{Finite Temperature Behavior of the Lattice Abelian Higgs Model}.


\bibitem{0901.4429}
\article[0901.4429]{C. Bonati, G. Cossu, A. D'Alessandro, M. D'Elia, A. Di Giacomo}{PoS}{LATTICE2008}{252}{2009}
{On the phase diagram of the Higgs SU(2) model}.


%


\bibitem{Abbott:1981re}
\article[Abbott:1981re]{L.F. Abbott, E. Farhi}{Phys. Lett.}{101B}{69}{1981}
{Are the Weak Interactions Strong?}.


\bibitem{hep-th/9812204}
\heparticle[hep-th/9812204]{G. 't Hooft}{Topological aspects of quantum chromodynamics}.


\bibitem{1608.05240}
\heparticle[1608.05240]{A. Trautner}{CP and other Symmetries of Symmetries}.


\bibitem{1505.07480}
\article[1505.07480]{C. Gross, O. Lebedev, Y. Mambrini}{JHEP}{1508}{158}{2015}
{Non-Abelian gauge fields as dark matter}.


\bibitem{1611.00365}
\article[1611.00365]{G. Arcadi, C. Gross, O. Lebedev, Y. Mambrini, S. Pokorski, T. Toma}{JHEP}{1612}{081}{2016}
{Multicomponent Dark Matter from Gauge Symmetry}.


\bibitem{1503.08749}
\article[1503.08749]{O. Antipin, M. Redi, A. Strumia, E. Vigiani}{JHEP}{1507}{039}{2015}
{Accidental Composite Dark Matter}.


\bibitem{1707.05380}
\article[1707.05380]{A. Mitridate, M. Redi, J. Smirnov, A. Strumia}{JHEP}{1710}{210}{2017}
{Dark Matter as a weakly coupled Dark Baryon}.


\bibitem{1811.06975}
\article[1811.06975]{R. Contino, A. Mitridate, A. Podo, M. Redi}{JHEP}{1902}{187}{2019}
{Gluequark Dark Matter}.


\bibitem{Gildener:1976ih}
\article[Gildener:1976ih]{E. Gildener, S. Weinberg}{Phys. Rev.}{D13}{3333}{1976}
{Symmetry Breaking and Scalar Bosons}.


\bibitem{hep-lat/9901004}
\article[hep-lat/9901004]{C.J. Morningstar, M.J. Peardon}{Phys. Rev.}{D60}{034509}{1999}
{The Glueball spectrum from an anisotropic lattice study}.


\bibitem{Dimopoulos:1980hn}
\article[Dimopoulos:1980hn]{S. Dimopoulos, S. Raby, L. Susskind}{Nucl. Phys.}{B173}{208}{1980}
{Light Composite Fermions}.


\bibitem{1607.05860}
\article[1607.05860]{A. Maas, P. T{\" o}rek}{Phys. Rev.}{D95}{014501}{2017}
{Predicting the singlet vector channel in a partially broken gauge-Higgs theory}.


\bibitem{1709.07477}
\article[1709.07477]{A. Maas, R. Sondenheimer, P. T{\" o}rek}{Annals Phys.}{402}{18}{2019-03}
{On the observable spectrum of theories with a Brout-Englert-Higgs effect}.


\bibitem{1804.04453}
\article[1804.04453]{A. Maas, P. T{\" o}rek}{Annals Phys.}{397}{303}{2018-10}
{The spectrum of an SU(3) gauge theory with a fundamental Higgs field}.


\bibitem{1606.00159}
\article[1606.00159]{K. Harigaya, M. Ibe, K. Kaneta, W. Nakano, M. Suzuki}{JHEP}{1608}{151}{2016}
{Thermal Relic Dark Matter Beyond the Unitarity Limit}.


\bibitem{1801.01135}
\article[1801.01135]{V. De Luca, A. Mitridate, M. Redi, J. Smirnov, A. Strumia}{Phys. Rev.}{D97}{115024}{2018}
{Colored Dark Matter}.


\bibitem{1811.08418}
\article[1811.08418]{C. Gross, A. Mitridate, M. Redi, J. Smirnov, A. Strumia}{Phys. Rev.}{D99}{016024}{2019}
{Cosmological Abundance of Colored Relics}.


\bibitem{Aprile:2018dbl}
\article[Aprile:2018dbl]{{\sc Xenon1T} Collaboration}{Phys. Rev. Lett.}{121}{111302}{2018}
{Dark Matter Search Results from a One Ton-Year Exposure of XENON1T}.


\bibitem{1503.02641}
\article[1503.02641]{{\sc Fermi LAT} Collaboration}{Phys. Rev. Lett.}{115}{231301}{2015}
{Searching for Dark Matter Annihilation from Milky Way Dwarf Spheroidal Galaxies with Six Years of Fermi Large Area Telescope Data}.


\bibitem{1505.05488}
\article[1505.05488]{D. Buttazzo, F. Sala, A. Tesi}{JHEP}{1511}{158}{2015}
{Singlet-like Higgs bosons at present and future colliders}.


\bibitem{1807.04743}
\article[1807.04743]{D. Buttazzo, D. Redigolo, F. Sala, A. Tesi}{JHEP}{1811}{144}{2018}
{Fusing Vectors into Scalars at High Energy Lepton Colliders}.


\bibitem{1607.08142}
\article[1607.08142]{{\sc HESS } Collaboration}{Phys. Rev. Lett.}{117}{111301}{2016}
{Search for dark matter annihilations towards the inner Galactic halo from 10 years of observations with H.E.S.S}.


\bibitem{1709.01483}
\article[1709.01483]{A. Morselli}{PoS}{ICRC2017}{921}{2017}
{The Dark Matter Programme of the Cherenkov Telescope Array}.


\bibitem{Witten:1983tx}
\article[Witten:1983tx]{E. Witten}{Nucl. Phys.}{B223}{433}{1983}
{Current Algebra, Baryons, and Quark Confinement}.


\bibitem{Graves}
\article{J.T. Graves}{Phil. Mag.}{26}{315}{1845}{\href{http://zs.thulb.uni-jena.de/receive/jportal_jparticle_00207304}{On a Connection between the General Theory of Normal Couples and the Theory of Complete Quadratic Functions of Two Variables}}.


\bibitem{hep-lat/0302023}
\article[hep-lat/0302023]{K. Holland, P. Minkowski, M. Pepe, U.J. Wiese}{Nucl. Phys.}{B668}{207}{2003}
{Exceptional confinement in $G_2$ gauge theory}.


\bibitem{1003.5912}
\article[1003.5912]{F. D'Eramo, J. Thaler}{JHEP}{1006}{109}{2010}
{Semi-annihilation of Dark Matter}.


\bibitem{Gunaydin:1973rs}
\article[Gunaydin:1973rs]{M. Gunaydin, F. Gursey}{J. Math. Phys.}{14}{1651}{1973}
{Quark structure and octonions}.


\end{thebibliography}
\end{document}